\documentclass[aip, jcp, reprint]{revtex4-2}
\pdfoutput=1
\usepackage{amsmath}
\usepackage{amssymb}
\usepackage{graphicx}
\usepackage{dcolumn}
\newcolumntype{d}[1]{D{.}{.}{#1}}
\usepackage{bm}
\usepackage[utf8]{inputenc}
\usepackage[T1]{fontenc}
\usepackage{mathptmx}

\newcolumntype{d}[1]{D{.}{.}{#1}}

\begin{document}

\title{From Order to Disorder of Alkanethiol SAMs on Complex Au (211), (221) and (311) Surfaces: Impact of the Substrate}

\author{Dimitrios Stefanakis}
\affiliation{Department of Materials Science \& Technology - University of Crete, Vassilika Voutes, 700~13 Heraklion, GREECE}
\email{dimstef@materials.uoc.gr}
\author{Vagelis Harmandaris}
\affiliation{Department of Mathematics \& Applied Mathematics - University of Crete, Vassilika Voutes, 700~13 Heraklion, GREECE}
\affiliation{Institute of Applied \& Computational Mathematics, Foundation for Research and Technology-Hellas, 711~10 Heraklion, GREECE}
\affiliation{Computation-Based Science and Technology Research Center, The Cyprus Institute, Nicosia 2121, CYPRUS}
\email{harman@uoc.gr}
\author{Georgios Kopidakis}
\affiliation{Department of Materials Science \& Technology - University of Crete, Vassilika Voutes, 700~13 Heraklion, GREECE}
\affiliation{Institute of Electronic Structure and Laser, Foundation for Research and Technology-Hellas, 711~10 Heraklion, GREECE}
\email{kopidaki@materials.uoc.gr}
\author{Ioannis Remediakis}
\affiliation{Department of Materials Science \& Technology - University of Crete, Vassilika Voutes, 700~13 Heraklion, GREECE}
\affiliation{Institute of Electronic Structure and Laser, Foundation for Research and Technology-Hellas, 711~10 Heraklion, GREECE}
\email{remed@materials.uoc.gr}

\keywords{Self Assembled Monolayer, Density Functional Theory, Molecular Dynamics, United Atom Model, Alkanethiols, Complex Au Surfaces, Au(111), Au(211), Au(221), Au(311), Atomistic Force Field, Atomistic Simulations}

\begin{abstract}
	We investigate the impact of the substrate on the structural properties and the morphology of alkanethiol self-assembled monolayers (SAMs) on gold, using first principles calculations and atomistic molecular dynamics simulations. We consider hexadecanethiols on Au(211), Au(221) and Au(311) surfaces which contain few-atom wide terraces separated by monoatomic steps similar to the complex Au surfaces used in experiments. The structure of the SAMs is probed via several structural properties including tilt angles, mean C atom heights from the surface, precession angles, gauche defects, gyration tensors and relative shape anisotropy. Comparing these properties to those of the well-studied SAMs on Au(111), we observe similarities but also striking differences. A clear order to disorder transition is observed by changing the substrate: well-ordered SAMs on (111) and (211) surfaces become mixed ordered-disordered structures on (311) and fully disordered on (221). The presence of steps on the Au surfaces also results in preferential tilt orientations with long-range order. Our results show that in addition to the expected grafting density dependence, the transition from order to disorder crucially depends on substrate morphology. The onset of ordering behavior is related to the atomic structure of the surface. The key parameter that affects long-range order is the energy for changing the dihedral angle between Au-S-$\mathrm{C^{(1)}-C^{(2)}}$ of the adsorbed alkanethiol.
\end{abstract}

\maketitle

\section{\label{sec:intro}Introduction}
Self-assembled monolayers (SAMs) are systems where (relatively small) molecules adsorbed on surfaces self-organize into, more or less, large ordered domains. For the self-assembly process both the molecule/surface and intermolecular interactions play an important role. Typically SAMs can be easily formed by spontaneous adsorption from gas or liquid phases, and this formation process is guided by a covalent linker: for clean metal substrates S is the most preferable one\cite{SCHREIBER2000151}. Alkanethiol monolayers on noble metal substrates (particularly Au or Ag) are the most common SAMs because of their continuously growing usage for promising applications: molecular biology, surface and materials science, inorganic chemistry, drug delivery and medical therapy\cite{doi:10.1021/ja710321g,doi:10.1002/anie.200904359}, surface functionalization\cite{doi:10.1002/mats.202000010}, catalysis and nanotechnology are some of the scientific fields that can benefit from these formations.

Due to the above reasons, several workers have studied experimentally\cite{doi:10.1063/1.1346676,doi:10.1021/ja036143d,doi:10.1021/la981374x,doi:10.1039/B505903H,SCHREIBER2000151,C4CP00596A} or theoretically\cite{doi:10.1021/jp067347u,doi:10.1021/jp200447k,doi:10.1021/jp0746289,doi:10.1021/la00013a028,doi:10.1021/la962055d,doi:10.1021/cr0300789} SAMs on Au(111) planar surfaces. On the other hand, very few studies have dealt with more interesting high-index surfaces\cite{LEUNG2018188, Nguyen_2012}. Surfaces with Miller indexes higher than one possess a periodic arrangement of terraces separated by infinitely long steps which sometimes are joined in kinks. For example, the (211) surface of a face-centered cubic (fcc) structured metal, such as gold, consists of three-atom wide close-packed terraces and monoatomic steps. While atoms on terraces have similar atomic environment as atoms on the flat (111) surface, step-edge atoms offer much stronger binding sites for alkanethiols\cite{doi:10.1063/1.4790368}. As a result, the overall properties of a SAM on Au(211) might be very different than those of a SAM on Au(111). Such a detailed study of the effects of surface structure on the properties of SAMs is missing. Moreover, a theoretical investigation for such systems is of great importance because such complex surfaces are also closely related to the surfaces of gold nanoparticles.

In the present study, we use a detailed and accurate all-atom classical force field that consists of interaction parameters that mainly come from previous works. Some parts of the potential are re-parameterized using first-principles calculations based on Density-Functional Theory (DFT). In particular, we derive a new interaction potential term for the $\mathrm{Au-S-C^{(1)}-C^{(2)}}$ dihedral angle, where Au is the nearest surface Au atom to the S and $\mathrm{C^{(1)}}$, $\mathrm{C^{(2)}}$ are the first and the second C atom in the chain. This was necessary to be done due to the lack of previous works concerning these complex surfaces. Then, we perform long time classical MD simulations, with the complete force field, to predict the structural properties of hexadecanethiols adsorbed on various Au complex surface. More specifically, we simulate SAMs on four different complex Au surfaces: (111), (211), (221) and (311). SAMs on Au(111) can be used as a reference system for the results of our own simulations as there is a lot of literature for them. The Au(211) surface has been selected because it has been shown that this is the surface that almost totally dominates thiolate-protected gold nanoparticles of diameters between \textasciitilde5 and \textasciitilde34~nm at the thermodynamic limit\cite{doi:10.1063/1.4790368}. The Au(311) surface has been selected as it has a nice coverage of fluorophore-labeled DNA and alkylthiol SAM on single crystal bead electrodes of Au\cite{LEUNG2018188} leading to biosensors construction. Moreover, this system has been studied among others by binding COOH-terminated alkanethiol molecules with AuNP surfaces which is useful and valuable for preparation of probe biomolecules for further biochip studies\cite{Nguyen_2012}. Finally, the Au(221) surface was selected because it has different step type, (111)/(111), compared to the (111)/(100) steps of (311) and (211). For all three complex surfaces, S atoms bind to Au atoms on the step edge without sharing of Au atoms. The distance between adjacent S atoms is therefore $2d$, where $d$ is the Au-Au distance in bulk Au, and the two lattice vectors are orthogonal, being parallel and perpendicular to the step edge. 

In addition to stepped surfaces, we perform calculations for ideally flat close-packed (111) with similar SAM-surface interaction as the stepped surfaces.  Several structures exist for SAMs on Au(111), the most studied one being the  $(\sqrt{3} \times \sqrt{3})R30^\circ$ with or without $( 4 \times 2)$ superstructure\cite{SCHREIBER2000151, doi:10.1021/cr0300789}. In its simplest form, the unit cell contains 3 surface Au atoms; S atoms of the SAM form a hexagonal lattice with S-S distance equal to $\sqrt{3} d$ and SAM grafting density $\frac{1}{3d^2}$ where $d$ is the Au-Au distance in bulk Au\cite{doi:10.1021/jp067347u}. Another known structure is the  $(2\times\sqrt{3})rect$ structure  which has 4 surface Au atoms per unit cell; S atoms of the SAM form an orthogonal lattice with S-S distances equal to $\sqrt{3}d$ and $2d$  with SAM grafting density $\frac{1}{2\sqrt{3}d^2}$\cite{doi:10.1063/1.4790368, doi:10.1063/1.1346676, doi:10.1021/ja036143d}. The differences in symmetry and grafting density result in different SAM properties. For example, the tilt angle is higher for the lower density SAM\cite{doi:10.1063/1.1346676,doi:10.1021/ja036143d}. A detailed study of SAMS on Au(111) can be found in several works in the literature and is beyond the scope of the present work. We use the less-common  $(2\times\sqrt{3})rect$ structure for (111) in order to have direct comparison between stepped and flat surfaces. In all four structures for Au we consider, S atoms of the SAM form orthogonal lattices with same S-S distance, and thus same linear grafting density along the two Au atoms that lie directly below the S atoms.

\begin{figure}
	\includegraphics[width=0.5\textwidth]{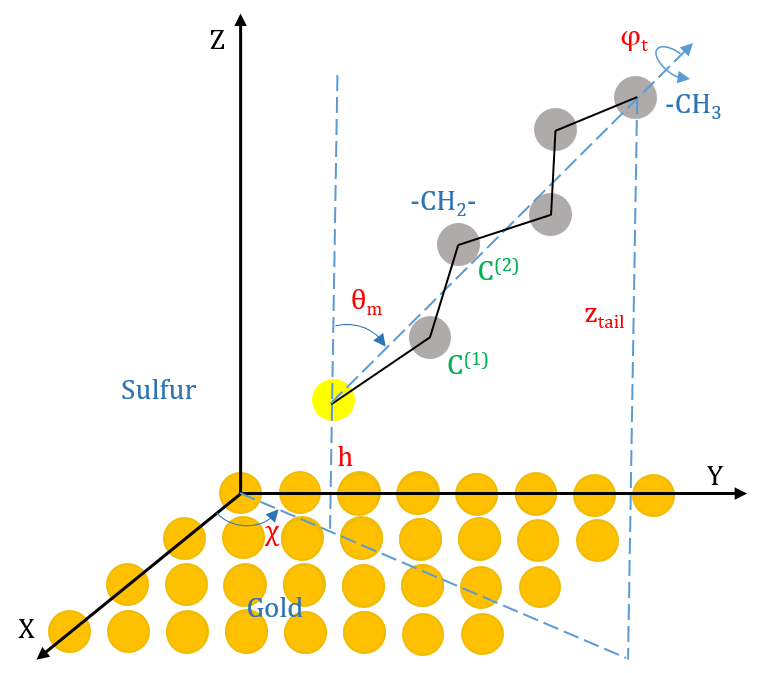}
	\caption{\label{fig:str_pr} Some structural properties of the studied systems: tilt angle ($\theta_m$), C atom distance from the slab ($z_{tail}$), the precession angle ($\chi$) and the torsion angle of the alkane chain ($\phi_t$).}
\end{figure}

For all above systems, we calculate a variety of structural parameters that can be defined in order to characterize them. Such properties include the tilt angle ($\theta_m$), the mean C atom distance according to its ranking along the chain ($h$) from the slab, the monolayer thickness ($z_{tail}$), the precession angle of the chain ($\chi$) and the percentage of Gauche defects of the alkane chain. 
The definitions of these quantities are shown schematicaly in the model of an alkanethiol on gold surface shown in Figure \ref{fig:str_pr}. The presence of Gauche defects (cis- instead of trans- or vice versa) is quite common for these molecules and will also be studied in detail here.

SAMs on planar surfaces are well-known to bind through defects, such as adatoms, vacancies and islands. When adatoms are present on a flat gold surface, S is found to bind to bridge site between adatom and surface-layer atom\cite{PhysRevLett.97.146103}. Similar local atomic arrangement is observed for thiol adsorption on Au clusters where again S binds to two under-coordinated Au atoms\cite{doi:10.1038/nchem.1352, Walter9157}. Instead of introducing defects on a perfect close-packed surface, we consider ideal surfaces with periodic arrangement of steps. Au atoms along the step edge are under-coordinated, having between five and seven neighbors while atoms on Au(111) have nine neighbors. In addition, atoms right next to these undercoordinated atoms have nine neighbors as they belong to (111) terraces. Therefore, the structures we consider have similar qualitative features as flat surfaces with adatoms while at the same time have perfect periodicity. These high-index surfaces contain a periodic arrangement of steps with various orientations and concentrations, and resemble model structures for defective planar surfaces.

Tilt angle, that is defined as the angle between the backbone of an alkanethiol and the normal to the substrate ($\theta_{m}$ in Figure \ref{fig:str_pr}), is a well studied property of various alkanethiol systems ($\mathrm{RS(CH_{2})_{n}CH_{3}}$) on metal substrates both theoretically\cite{doi:10.1021/la962055d,doi:10.1021/la00013a028,doi:10.1021/cr0300789,doi:10.1039/B907301A,doi:10.1039/B505903H,doi:10.1021/jp0746289,doi:10.1038/nchem.1352} and experimentally\cite{doi:10.1063/1.1346676,doi:10.1021/ja036143d}. Most of these studies refer to close packed arrangements of molecules with the $(\sqrt{3}\times\sqrt{3})R30^{o}$ hexagonal periodicity relative to a Au substrate or to a secondary $c(4\times2)$ superstructure on it; the value of tilt angles on such arrangements varies between \textasciitilde$30^{o}$ and \textasciitilde$35^{o}$ at room temperature for various values of $n$. However, a few cases with a less dense arrangement of $(2\times\sqrt{3})rect$ have been observed experimentally\cite{doi:10.1063/1.1346676,doi:10.1021/ja036143d} as well as theoretically\cite{doi:10.1021/jp200447k}, where the tilt angle differs a lot from the above as its value lies around $50^{o}$.

Such formations are observed experimentally as metastable states that are quickly transformed into $(\sqrt{3}\times\sqrt{3})R30^{o}$ arrangements after some disturbance, while in theoretical studies, where the bond distance between the alkanethiol chains was kept fixed, the tilt angle remained unchanged at \textasciitilde$50^{o}$. Adding more C atoms in the chain, this flexible structure seems to become more stable and can be observed in simulations where the distances between chains are flexible\cite{doi:10.1021/jp200447k}.

We should also note that a well-studied structural characteristic of the SAMs is the so called "odd-even" effect\cite{B004232N,doi:10.1021/acs.jpcc.5b07899,doi:10.1021/acs.jpcc.9b05299,doi:10.1021/cr050258d}. According to this, alterations of the properties of SAMs structures depending on the odd or even number of the C atoms in the alkane chain have been observed for chain lengths between 2 and 18. Especially for SAMs on Au(111) investigations, one important property that is affected by this effect is the tilt angle of the alkane chain which tends to be larger for odd numbers of C atoms in chain than for even numbers. Here we consider SAMs with a constant alkane length of 16 C atoms.

The binding site of S for flat Au surface can be bridge, hollow or on-top, depending on the alkanethiol length and surface defects\cite{doi:10.1021/jp100522n}. DFT calculations for the same surfaces used in the present study show very strong preference for adsorption on the bridge site. For example, in (211) surface, adsorption on bridge site has lower adsorption energy by more than 0.5 eV per molecule compared to the top site\cite{doi:10.1063/1.4790368}.

\section{\label{sec:meth}Model and Simulation Methodology}
\subsection{\label{sec:meth1}Sample preparation and construction}
{\em Unit cell generation:} The construction of our samples, was based on the results of Barmparis et al.\cite{doi:10.1063/1.4790368} In that work, the authors had considered methanethiolates ($\mathrm{RS-}\ \textrm{with}\ \mathrm{R=CH_{3}}$) adsorbed on various Au($hkl$) surfaces. Using Density Functional Theory (DFT) simulations, they considered every possible adsorption geometry on all Au($hkl$) with indices $h,k,l<4$. Gold surfaces were modeled using slabs of Au with periodic boundary conditions in directions along the surface plane. We use the minimum energy structure of methanethiolate for each one of the four Au surfaces considered in the present study. Starting from this minimum adsorption energy state of each mentioned surface, we developed new structures in sp$^3$ order via a geometric procedure as follows:
\begin{enumerate}
	\item Substitution of one H atom of the methyl group with one methylene group ($\mathrm{-CH_2-}$).
	\item On the free bond of this methylene another methylene was added.
	\item The above step was repeated until we reach the desired number of C atoms in the chain. The last added group was a methyl group.
\end{enumerate}
This way, we were able to construct the alkanethiol chains we needed consisting of sixteen C atoms, C$_{16}$. Distances and angles of the bonds were fixed initially to values known from the literature (bond distance of C\nobreakdash-C = 1.54 \AA, angles of H\nobreakdash-C\nobreakdash-H, C\nobreakdash-C\nobreakdash-C = $\mathrm{109.47^o}$), while the distances between C and H atoms are given by the initial sample. The thickness of the Au slab was at least 8 \AA\ while the lattice constant for all surfaces was 4.22 \AA. This value, which is close to the experimental one (4.08 \AA), was used by Barmparis et al.\cite{doi:10.1063/1.4790368} and is preserved in our calculations for compatibility reasons.

\begin{figure*}
	\includegraphics[width=1.0\textwidth]{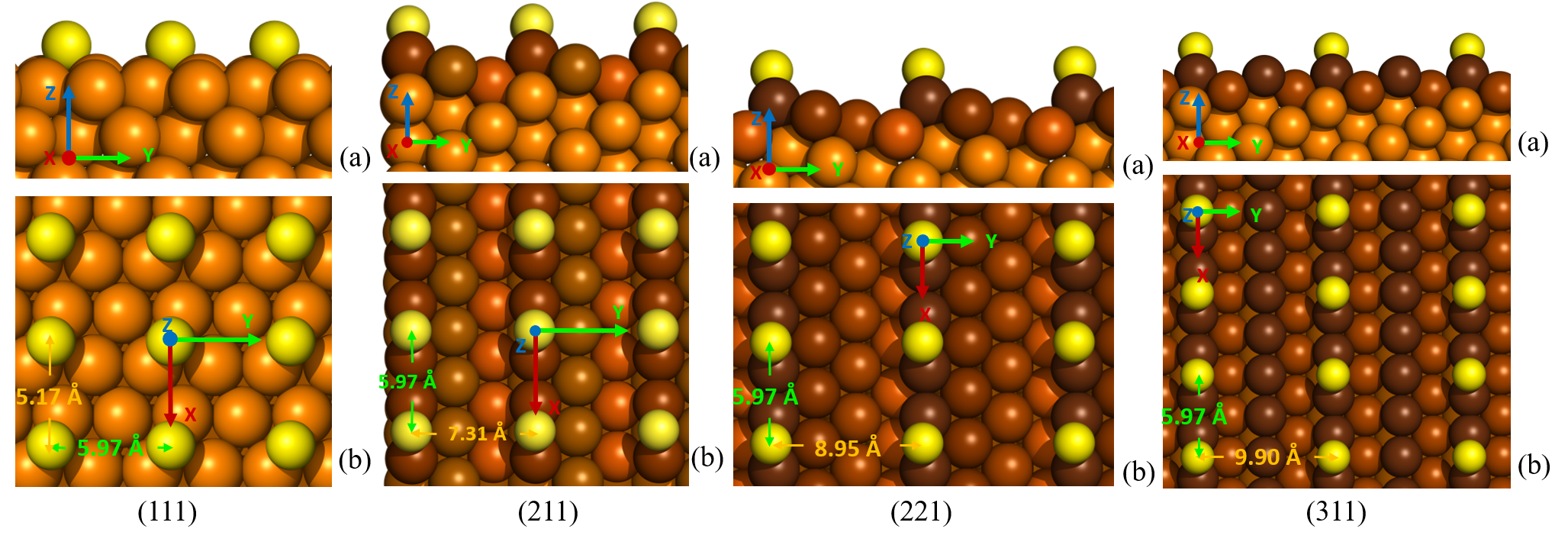}
	\caption{\label{fig:Au_S_multi_total} Geometrical characteristics for the (111), (211), (221) and (311) surfaces and the S positions on them: (a) Side view, (b) Top view. The color shade of the Au atoms indicates proximity to the surface, the darkest ones being those of the outermost layers.}
\end{figure*}

Characteristics of the various surfaces are summarized in Table \ref{tab:char} and are demonstrated in Figure \ref{fig:Au_S_multi_total}. In this figure, we used different shades of gold to show the distance of Au atoms from the surface. Step-edge atoms, shown with darkest color in Figure 2, are the ones that are bonded to S atoms of the alkanethiol. Each S atom is bonded to two step-edge Au atoms, and each step-edge Au atom is bonded to one S atom. The position of S in the middle of the bridge site is dictated by the DFT calculations\cite{doi:10.1063/1.4790368} which were used as a starting point for the present work. In that work, several initial positions of S were considered, and each structure was fully relaxed to find the lowest-energy configuration. Middle of the bridge site was the preferred adsorption geometry for S for all three stepped surfaces considered here. The main features of the different Au surfaces modeled in this work are discussed below.

\begin{table*}
	\caption{\label{tab:char}Characteristics of the various surfaces}
	\centering
	\resizebox{\textwidth}{!}{%
		\begin{tabular}{lcccc}
			&\multicolumn{4}{c}{Surfaces}\\
			\hline
			&Au(111)&Au(211)&Au(221)&Au(311)\\
			\hline
			Surface dimensions of a single cell (nm$^2$)&0.597~$\times$~0.517&0.597~$\times$~0.731&0.597~$\times$~0.895&0.597~$\times$~0.990\\
			Surface area of a single cell (nm$^2$)&0.309&0.436&0.534&0.591\\
			Grafting density (nm$^{-2}$)&3.24&2.29&1.87&1.69\\
			Total surface dimensions (nm)&17.9$\times$15.5&17.9$\times$21.9&17.9$\times$26.9&17.9$\times$29.7\\
			Total slab surface (nm$^2$)&280.80&394.20&486.00&534.60\\
			Number of Au atoms&25200&19800&30600&43200\\
			Microfacet notation&\ldots&3(111)$\times$(100)&4(111)$\times$(111)&2(111)$\times$(100)\\
		\end{tabular}
	}
\end{table*}

In the stepped surfaces we consider, the grafting density is dictated by the DFT simulations that show strong preference for step-edge binding of S on stepped surfaces. The grafting density is $\frac{1}{ndL}$ where $d$ is the distance between neighboring Au atoms along the step and $L$ the distance between steps. In our simulation, we place one S atom every second Au atom, therefore $n=2$. As shown in Table 1, the grafting densities we consider are between 1.7 and 3.2 nm$^{-2}$. Experimental grafting densities for alkanethiols range from 0.2 nm$^{-2}$\cite{doi:10.1021/la00081a009} to 4.6 nm$^{-2}$\cite{doi:10.1021/jp067347u}. Below we give detailed information for the characteristics of each surface.

\paragraph{\label{meth11}The Au(111) surface:}
In Au(111), S atoms are on a bridge site between two Au atoms of the surface and are arranged in a rectangular lattice with dimensions of 0.597~$\times$~0.517~nm$^2$. This gives a grafting density of 3.24~nm$^{-2}$. This structure, described as $(2\times\sqrt{3})rect$, although observed experimentally\cite{	doi:10.1063/1.1346676}, has a bit  lower density than the most common SAM structure for Au(111) which is $(\sqrt{3}\times\sqrt{3})R30^\circ$.  We chose to use $(2\times\sqrt{3})rect$ for the perfectly flat Au(111) in order to have four SAM structures with identical arrangement of S atoms both in terms of symmetry (S atoms form rectangular lattices) and S-S distance. With this choice, symmetry and S-S distance is the same in all four systems we consider. The S-S distance is  twice the nearest neighbor distance of bulk Au.

\paragraph{\label{meth12}The Au(211) surface:}
This surface consists of 3-atoms wide terraces and an 1-atom step. On the terraces, atoms have the same atomic configuration as in (111) surface, while on the step atoms resemble the structure of (100). The microfacet notation\cite{VANHOVE1980489} for this surface is therefore 3(111)$\times$(100) as shown in Table \ref{tab:char}. The S atoms are positioned on a bridge site between two Au atoms over the edge of the steps and are arranged in a rectangular lattice with dimensions of 0.597~$\times$~0.731~nm$^2$. This gives a grafting density of 2.29~nm$^{-2}$.

\paragraph{\label{meth13}The Au(221) surface:}
This surface is similar to the Au(211) since it consists of 4-atoms wide terraces and an 1-atom step. On terraces atoms have the (111) configuration, therefore the microfacet notation is 4(111)$\times$(111). The S atoms are positioned on a bridge site between two Au atoms over the edge of the steps and are arranged in a rectangular lattice with dimensions of 0.597~$\times$~0.895~nm$^2$. This gives a grafting density of 1.87~nm$^{-2}$.

\paragraph{\label{meth14}The Au(311) surface:}
The Au(311) surface consists of 2-atoms wide terraces and an 1-atom step. The structure of (311) is similar to that of (211), the only difference being the 2-atom wide terraces compared to 3-atom-wide terraces in (211). Thus, the microfacet notation for this surface is 2(111)$\times$(100). The S atoms are positioned on a bridge site between two Au atoms over the edge of every other step and are arranged in a rectangular lattice with dimensions of 0.597~$\times$~0.990~nm$^2$. This gives a grafting density of 1.69~nm$^{-2}$.

{\em Final sample construction:} The final sample for each kind of surface was formed by repeating the above initial cells 30 times on both x- and y- axes providing SAMs with 900 alkanethioles. 

\subsection{\label{sec:meth3} Atomistic force field (interaction potentials)}
The entire force field used in the present work is based on previous classical force fields for SAMs~\cite{doi:10.1021/jp067347u,doi:10.1021/jp0746289} that is extended, as described below.

The interatomic potentials used here are described in Table \ref{tab:potentials}. The majority of these potentials were taken from the literature. Here, we consider immobile alkanethiols, where the S-Au bond stays fixed throughout the simulation. This is a reasonable approximation, given that binding of S to step-edge atoms is extremely strong, and it is not likely that the S-Au bond can break at room temperature. Barmparis et.al\cite{doi:10.1063/1.4790368} found that the lowest values of the alkanethiols adsorption energies over the above surfaces are -0.146, -0.81, -0.68 and -0.75 eV for the (111), (211), (221) and (311) surfaces respectively which are far from the typical kinetic energy of a gas molecule ($\frac{3}{2}kT,\ k:$ Boltzmann's constant) at \textasciitilde$3.9\times10^{-2}$ eV. The positions of Au atoms have been kept frozen as well for the same reason.

For the rest of our particles, the total potential energy ($V_{total}$) of a particle is\\
\begin{equation}
V_{total}=V_{b}+V_{nb}
\label{eq:Vtot}
\end{equation}
where $V_{b}$ stands for the intramolecular (bonded) and $V_{nb}$ for the intermolecular (non-bonded) interactions. The $V_{b}$ and $V_{nb}$ are given by\\
\begin{subequations}
	\label{Vpart}
	\begin{eqnarray}
	V_{b}&=&V_{stretch}+V_{bend}+V_{tor}\\
	V_{nb}&=&V_{LJ}(r)=4\epsilon_{ij}\biggl(\Bigl(\frac{\sigma_{ij}}{r_{ij}}\Bigr)^{12}-\Bigl(\frac{\sigma_{ij}}{r_{ij}}\Bigr)^{6}\biggr),
	\end{eqnarray}
\end{subequations}
respectively, and described in Table \ref{tab:potentials}.

The bond-stretching ($V_{stretch}$), the bond-bending ($V_{bend}$) and the dihedral angles interactions ($V_{tor}$) of the SAMs chains were taken from the literature~~\cite{doi:10.1021/jp067347u,doi:10.1021/jp0746289}. 
On the contrary, due to the lack of a detailed interaction potential for the $\mathrm{Au-S-CH_2-CH_2}$ dihedral angles we have developed a new one. For that we've performed new DFT calculations and parametrized a polynomial function for each surface, as it is presented in the Section "Calculation of Au-S-C$^{(1)}$-C$^{(2)}$ dihedral angle potentials"\ref{sec:meth31}.

The non-bonded interactions of $\mathrm{S-CH_x}\ (x=2,3)$, $\mathrm{CH_x-CH_y}\ (x,y=2,3)$ and $\mathrm{Au-CH_x}\ (x=2,3)$ were described by the typical 12-6 Lennard-Jones potential of Equation \ref{Vpart}b. The estimation of the proper values for $\epsilon_{ij}$ and $\sigma_{ij}$ in LJ interactions were based on Lorentz-Berthelot rules ($\sigma_{ij}=\frac{1}{2}(\sigma_{ii}+\sigma_{jj})$ and $\epsilon_{ij}=\sqrt{\vphantom{b}\epsilon_{ii}\epsilon_{jj}}$) using the values presented in Table \ref{tab:potentials}.

\begin{table*}
	\caption{\label{tab:potentials}Interaction parameters of the molecular models used in simulations}
	\centering
	\resizebox{\textwidth}{!}{%
		\begin{tabular}{lcccc}
			Type of interaction and potential function&\multicolumn{4}{c}{Type of interacting sites}\\
			\hline
			\multicolumn{5}{l}{Bond-stretching interactions}\\
			$V_{stretch}(r)=\frac{1}{2}k_{s}(r-r_{0})^2$&$\mathrm{CH_2-CH_x}$ (x=2,3)&$\mathrm{S-CH_2}$&$\mathrm{Au-S}$&$\mathrm{Au-Au}$ (slab)\\
			$r_{0}\ (nm)$&0.154&0.181&&\\
			$k_{s}\ (\frac{kJ}{mol} \times nm^{-2})$&217568.00\cite{doi:10.1021/jp0746289}&185769.00\cite{doi:10.1021/jp0746289}&frozen&frozen\\
			\hline
			\multicolumn{5}{l}{Bond-bending interactions}\\
			$V_{bend}(\theta)=\frac{1}{2}k_{b}(\theta-\theta_{0})^2$&$\mathrm{CH_2-CH_2-CH_x}$ (x=2,3)&$\mathrm{S-CH_2-CH_2}$&$\mathrm{Au-S-CH_2}$&$\mathrm{Au-Au-Au}$ (slab)\\
			$\theta_{0}\ (deg)$&109.5&114.4\cite{doi:10.1021/jp067347u}&110.1\cite{doi:10.1063/1.4790368}&\\
			$k_{b}\ (\frac{kJ}{mol} \times rad^{-2})$&519.653&519.653&519.653\cite{doi:10.1021/jp067347u}&frozen\\
			\hline
			\multicolumn{5}{l}{Dihedral angle interactions}\\
			$V_{tor}(\phi)=\sum\limits_{i=0}^{5}\alpha_{i}\cos(\phi)$&\multicolumn{2}{c}{$\mathrm{CH_2-CH_2-CH_2-CH_x}$}&$\mathrm{S-CH_2-CH_2-CH_2}$&$\mathrm{Au-Au-Au-Au}$ (slab)\\
			(Ryckaert - Bellemans function)&\multicolumn{2}{c}{(x=2,3)}&&\\
			$\alpha_{i}\ (\frac{kJ}{mol})$&\multicolumn{3}{c}{$\alpha_0=9.2759\ /\ \alpha_1=12.1545\ /\ \alpha_2=-13.1168$}&frozen\\
			&\multicolumn{3}{c}{$\alpha_3=-3.0585\ /\ \alpha_4=26.2378\ /\ \alpha_5=-31.4929$\cite{doi:10.1021/jp067347u}}&\\
			\hline
			\multicolumn{2}{l}{Dihedral interactions for the $\mathrm{Au-S-CH_2-CH_2}$ angle}&\multicolumn{3}{c}{Au surfaces}\\
			\multicolumn{1}{c}{Polynomial coefficients}&(111)&(211)&(221)&(311)\\
			\multicolumn{1}{c}{$a_{0}$}&$2.063716\times10^{1}$&$1.938784\times10^{1}$&$14.79002\times10^{1}$&$1.856027\times10^{1}$\\
			\multicolumn{1}{c}{$a_{1}$}&$-1.415672\times10^{-1}$&$-1.316188\times10^{-1}$&$1.697293\times10^{-1}$&$2.507222\times10^{-1}$\\
			\multicolumn{1}{c}{$a_{2}$}&$-2.655100\times10^{-3}$&$-3.480334\times10^{-3}$&$-2.947214\times10^{-3}$&$-2.147820\times10^{-3}$\\
			\multicolumn{1}{c}{$a_{3}$}&$-1.259557\times10^{-05}$&$-4.817540\times10^{-6}$&$-5.552760\times10^{-6}$&$-1.823129\times10^{-5}$\\
			\multicolumn{1}{c}{$a_{4}$}&$2.743923\times10^{-07}$&$5.119324\times10^{-7}$&$5.090070\times10^{-7}$&$2.218595\times10^{-7}$\\
			\multicolumn{1}{c}{$a_{5}$}&$3.160848\times10^{-09}$&$1.715997\times10^{-9}$&$-1.184259\times10^{-9}$&$-3.324587\times10^{-10}$\\
			\multicolumn{1}{c}{$a_{6}$}&$-1.457525\times10^{-11}$&$-3.462513\times10^{-11}$&$-3.632844\times10^{-11}$&$-1.365227\times10^{-11}$\\
			\multicolumn{1}{c}{$a_{7}$}&$-1.864453\times10^{-13}$&$-7.556656\times10^{-14}$&$8.703494\times10^{-14}$&$5.466089\times10^{-14}$\\
			\multicolumn{1}{c}{$a_{8}$}&$3.337915\times10^{-16}$&$1.006355\times10^{-15}$&$1.076216\times10^{-15}$&$3.809166\times10^{-16}$\\
			\multicolumn{1}{c}{$a_{9}$}&$4.553404\times10^{-18}$&$9.766564\times10^{-19}$&$-2.113789\times10^{-18}$&$-1.473163\times10^{-18}$\\
			\multicolumn{1}{c}{$a_{10}$}&$-2.656180\times10^{-21}$&$-1.045875\times10^{-20}$&$-1.137111\times10^{-20}$&$-3.823418\times10^{-21}$\\
			\multicolumn{1}{c}{$a_{11}$}&$-4.046784\times10^{-23}$&$-5.492738\times10^{-25}$&$1.743463\times10^{-23}$&$1.269500\times10^{-23}$\\
			\multicolumn{1}{c}{$a_{12}$}&&&$2.807201\times10^{-27}$&\\
			\hline
			\multicolumn{5}{l}{Non-bonded interactions}\\
			$V_{LJ}(r)=4\epsilon_{ij}\Bigl(\bigl(\frac{\sigma_{ij}}{r_{ij}}\bigr)^{12}-\bigl(\frac{\sigma_{ij}}{r_{ij}}\bigr)^{6}\Bigr)$&$\mathrm{S}$&$\mathrm{CH_2}$&$\mathrm{CH_3}$&$\mathrm{Au}$\\
			$\epsilon_{ij}\ (\frac{kJ}{mol})$&1.6628&0.4937&0.7326&0.1632\\
			$\sigma_{ij}\ (nm)$&0.4250&0.3905&0.3905&0.2935\cite{doi:10.1021/jp0746289}\\
		\end{tabular}
	}
\end{table*}

Another interesting potential function is the one that describes the energy cost related to the Au-S-C angle, with the surface of the substrate considered as a plane. This potential, although not used in the present calculation, has a large importance in the context of so called odd-even effects for SAMs. The potential is of harmonic type, $V(\theta)=\frac{1}{2}k_b(\theta-\theta_0)^2$, with $k_b$ the bond-bending constant given in Table 2. The angle $\theta_0$ equals 123.1$^\circ$ for (111), 113.3$^\circ$  for (211),  108.1$^\circ$ for (221) and 111.5$^\circ$  for (311), respectively.

\subsection{\label{sec:meth31}Calculation of Au-S-C$^{(1)}$-C$^{(2)}$ dihedral angle potentials}
\begin{figure*}
	\includegraphics[width=1.0\textwidth]{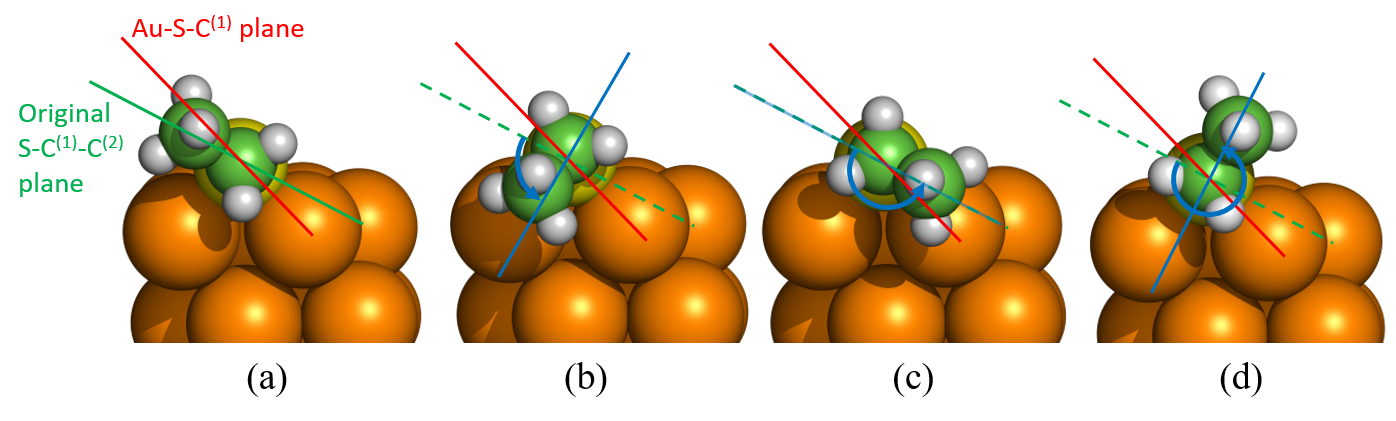}
	\caption{\label{fig:dih_pr} Calculation process of the potential for the Au-S-$\mathrm{C^{(1)}-C^{(2)}}$ dihedral on a Au(211) surface. (a), (b), (c) and (d) show the dihedral planes at the initial (original) position and after the rotation at 90, 180 and 270 degrees respectively. \emph{Red:} the Au-S-C plane, \emph{Green}: the S-C-C plane, \emph{Blue}: the new S-$\mathrm{C^{(1)}-C^{(2)}}$ plane after rotation. Note the slight displacement of the C atoms due to the slab repulsion when they seem to approach it in (b) and (c) where the $\mathrm{C^{(1)}}$ atom has been moved slightly up and left in comparison to its initial position.}
\end{figure*}

The atomistic force field described above, as well as the majority of the parameter values in Table 2, originated from previous works for flat Au surfaces where the dihedral $\mathrm{Au-S-C^{(1)}-C^{(2)}}$ does not play important role as it does for stepped surfaces.  The complexity of our surfaces and the lack of previous calculations for the potential of Au\nobreakdash-S\nobreakdash-$\mathrm{C^{(1)}-C^{(2)}}$ dihedral angles, guided us to make a new calculation method for it. Using state-of-the-art electronic structure methods, we calculate the energy of adsorbed ethanethiol at different (fixed) values of the dihedral angle, $\phi$, and fit the results to an analytical function of $\phi$. In this way, we end up with an accurate potential that takes into account variations of dihedral angle in adsorbed alkanethiols. The idea is to build structures containing ethanethiols over each of the mentioned surfaces, using the method already described in Section "Sample preparation and construction"\ref{sec:meth1} and calculate the potential for a number of angles by rotating the S\nobreakdash-$\mathrm{C^{(1)}}$ bond by 10 degrees at a time, starting from the original position. This process is shown in Figure \ref{fig:dih_pr} for the (211) surface; identical processes are used for the rest of the mentioned surfaces. In order to make the Au\nobreakdash-S\nobreakdash-$\mathrm{C^{(1)}-C^{(2)}}$ chain, we consider the Au atom with the smallest distance from the S atom of the ethanethiol; note that the S atom is positioned in the middle of a bridge site between two Au atoms thus the two distances were almost equal; see also the schematic representation in Figure \ref{fig:Au_S_multi_total}.

The calculations were performed according to the Kohn–Sham's\cite{PhysRev.140.A1133} approach of Hohenberg– Kohn's DFT\cite{PhysRev.136.B864} theory by using the ASE's\cite{ase} GPAW\cite{gpaw} code at Finite Difference mode. The space grid points were set to have a distance of 0.2 \AA\ between them, while the k-points were set to 2, 2, 1 for the x-, y- and z- axes respectively. The exchange correlation functional was the revised Perdew-Burke-Ernzerhof (RPBE)\cite{PhysRevB.59.7413}. The systems were relaxed until they reached the lowest energy which was finally selected. In some angle sites, where the second C atom seemed to enter between the atoms of the slab surface, the system was very unstable. This caused large variations in energy during relaxation process and some displacement from their expected positions was observed (Figure \ref{fig:dih_pr} b-c). Thus, in these situations the systems never converged and, as a result, we selected the lowest energy during a long relaxation process.

Starting from these data, we fitted a polynomial function for the potential difference between the calculated value and the lowest calculated value with respect to the Au\nobreakdash-S\nobreakdash-$\mathrm{C^{(1)}-C^{(2)}}$ dihedral angle $\phi_{rot}$ on each set of them in order to have this part of the potential scheme. Due to the periodicity of the potential, in the calculation of polynomials we have ensured that the value, as well as their first and second derivatives at the initial and the final angle of calculation, are respectively equal. This was achieved with accuracy between $10^{-10}$ and $10^{-7}$ depending on the examined surface. The polynomials found to be of 11th (for surfaces (111), (211) and (311)) and 12th (for 221 surface) grade and the results are demonstrated in Table \ref{tab:potentials}. The dihedral angle was finally fixed so that $\phi_{rot}^{(cis)}=0$, according to the \mbox{IUPAC/IUB} convention.

\begin{figure*}
	\includegraphics[width=1.0\textwidth]{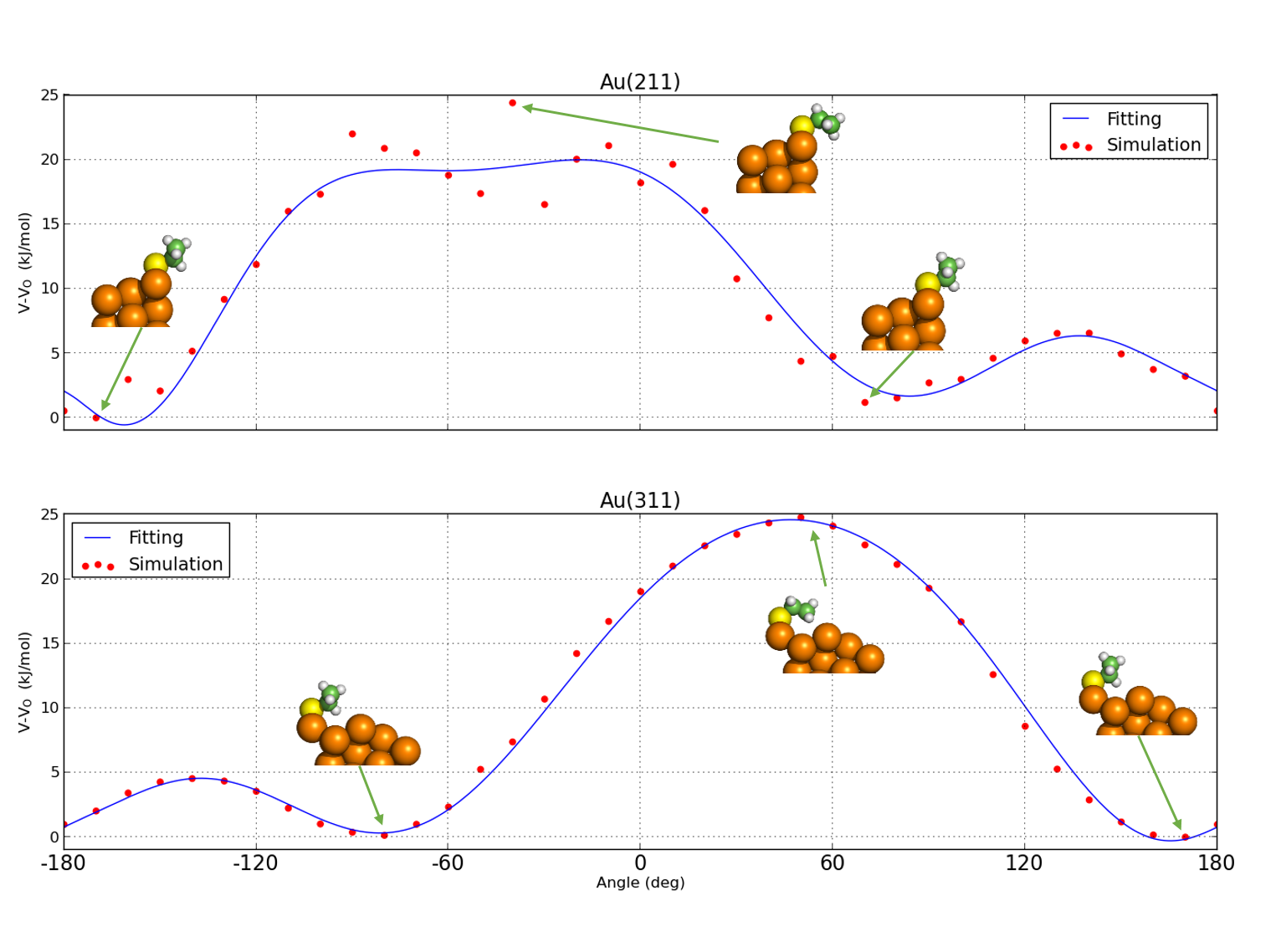}
	\caption{\label{fig:pot_dih_struct} Potential vs. the $\mathrm{Au-S-C^{(1)}-C^{(2)}}$ dihedral angle on Au surfaces. The fitting is not so good in sites where there was strong repulsion to C atoms from the slab atoms. The images demonstrate the positions of atoms in selected dihedral angles. Angles with respect to $\phi_{rot}^{(cis)}=0$, according the \mbox{IUPAC/IUB} convention.}
\end{figure*}

We plotted the simulation data (obtained by the DFT calculations), in conjuction with the fitting values as shown in Figure \ref{fig:pot_dih_struct} for Au(211) and Au(311). The potential on the vertical axis is presented in $kJ/mol$ (more specifically this is the difference between the original value $V$ and the calculated lowest value $V_{0}$), vs. the angle in degrees on the horizontal axis. The diagrams are shifted properly in order to be plotted according to the \mbox{IUPAC/IUB} ($\phi_{cis}=0$).

Although the potential functions for the (111) and (311) surfaces fit very well the calculated values, there is a higher deviation of values for the accuracy in the (211) and (221) configurations especially at angles where the second C atom seems to ``penetrate'' into the slab. The reason of course is that these are non permitted sites because the energy of the system is very high there. However, in spite of the deviation from the expected accuracy, we consider that these potential functions fit quite well the purpose they were constructed for.

Comparing the new dihedral angle potentials for the Au-S-C-C angle to other well-known potential for dihedral angles, we observe several similarities and differences. The potential of Ghorai and Glotzer \cite{doi:10.1021/jp0746289} ,  for the S-C-C-C dihedral angle, which was originaly parametrized for the C-C-C-C dihedral angle, is a third degree polynomial of $(1+\cos\phi)$, where $\phi$ is the dihedral angle. The two potentials are smooth periodic functions of $\phi$, with more than one minimum. The maximum energy is near $\phi=0$ and the minimum near $\phi=\pi$.  The energy difference between minimum and maximum energy is 35 kJ/mol for  Ghorai and Glotzer potential and 25 kJ/mol for the present potential We thus find a softer potential compared to the one for alkanes, which is a result of the presence of a metal atom in our case. The  Ghorai and Glotzer potential has two local minima located at angles $\approx \pm 60^\circ$. We only find one such local minimum for each surface as the presence of the Au atoms makes it energetically unfavorable for the second methyl group to be located close to the surface. Our potential has a large plateau region of high energy values between zero and 60 degrees; these conformations correspond to CH$_n$ groups being very close to Au atoms (see Fig. \ref{fig:pot_dih_struct}).

\subsection{\label{sec:meth4}Atomistic simulations}
After identifying the structures and the potential energy functions, we proceeded to perform the Molecular Dynamics (MD) simulations, using the GROMACS\cite{gromacs} open source package.

The MD simulations were performed using the NVT ensemble (canonical ensemble), and the Nos\'e-Hoover thermostat for temperature coupling on the whole system ($\tau$=0.2ps) at T=300K\cite{doi:10.1080/00268978400101201, PhysRevA.31.1695}. We used a large simulation box in z direction with more than 1 nm of vacuum above the SAMs and $30\times 30=900$ thiol molecules. Due to the presence of the vacuum region, the system can arrange its structure to reach the equilibrium density.  The $x-$ and $y-$ periodicity cannot be modified due to the presence of the thick Au slab underneath the SAM. At the conditions of the present study, the lattice constant of Au is not expected to change with pressure or temperature, so there is no need to use NPT ensemble. We also used the Verlet algorithm (Leap-Frog approximation) for the integration of the equations of motion\cite{doi:10.1063/1.442716}. Periodic boundary conditions were applied in all three dimensions.

Especially for z-axis, the interval between the slabs was kept over 5 nm in order to prevent our systems from vertical interactions. All structures were gradually exposed to temperatures of 500, 400 and 350 Kelvin; the output of the latter being the configuration that was eventually studied at 300K. The simulation time was 200ns for the (111), (211), (221) surfaces and 400ns for (311) to ensure proper equilibration and accurate calculation of the structural properties.

The question of ergodicity (time average equals ensemble average) is always important in simulations of self-assembled systems. To check that our simulation respects this principle, we have performed multiple MD runs (from 3 to 5) for each system in order to ensure that they end up at similar states, especially for the (221) and (311) ones that show less or no order at their final states. We tried several different initial conditions and also  performed runs at high temperature and then cool down at room temperature. In all cases, the key features of the final states were the same and independent of the initial state or the equilibration method.

\section{\label{sec:results} Results}
\subsection{\label{sec:sys_equil}Systems equilibrium and tilt angles}
In order to estimate the equilibrium state of each examined system, we observed the time evolution of the tilt angle for the C$_{16}$S chains on every surface. The time of convergence was 19, 17 and 260 ns for the (111), (211) and (311) systems, respectively, until they reached an ordered state. The (221) system was converged at 33 ns but it had never been able to reach an ordered state. Tilt angles and other structural parameters are tabulated in Table \ref{tab:results}. These are the average values from the analysis of the accumulated configurations, after equilibration was reached.

\begin{figure}
	\includegraphics[width=0.5\textwidth]{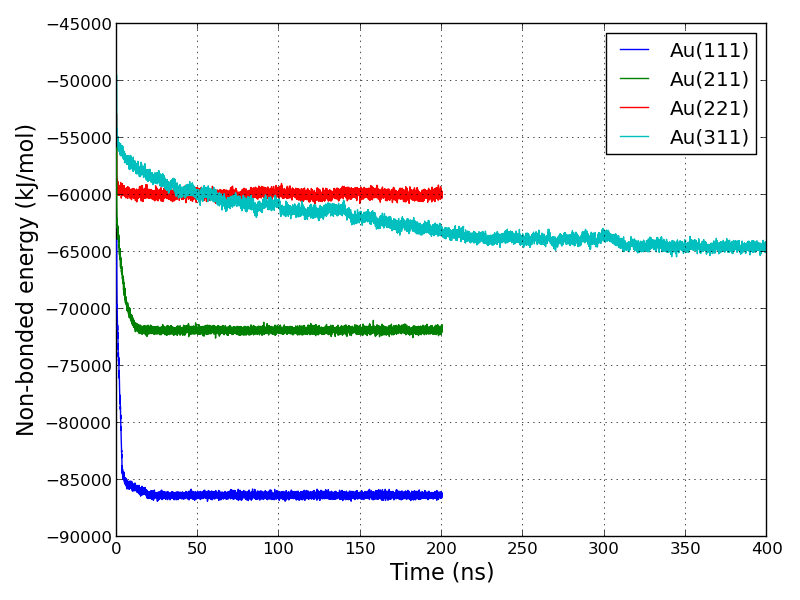}
	\caption{\label{fig:energy_evolution} Time evolution of the non-bonded potential energy for C$_{16}$S chains on Au surfaces. The energy of the (ordered) SAMs on the (111) and (211), and the (amorphous) one on (221), converge within about 50ns, whereas the one on the (311), which includes both ordered and amorphous domains, converges in much longer time scale, of about 300ns, compared to the rest.}
\end{figure}

\begin{table*}
	\caption{\label{tab:results}Selected results from the MD simulations}
	\centering
	\resizebox{\textwidth}{!}{%
		\begin{tabular}{lcccc}
			&\multicolumn{4}{c}{Surfaces}\\
			\hline
			Properties&(111)&(211)&(221)&(311)\\
			\hline
			Time of structural properties convergence (ns)&19&17&33&260\\
			Tilt angle <$\theta_{m}$> (deg)&52.6$\pm$2.8&61.1$\pm$3.1&61.3$\pm$16.7&69.6$\pm$3.2\footnotemark[1]\\
			Mean height of last C atom <$z_{tail}$>(nm)&1.44$\pm$0.06&1.12$\pm$0.07&0.84$\pm$0.38&0.96$\pm$0.22\\
			Precession angle <$\chi$> (deg)&147.2$\pm$2.7&235.4$\pm$4.3&&43.9$\pm$2.6\footnotemark[1]\\
			Gauche defects of the last methyl in chain&8.8\%&20.6\%&&18.2\%\\
			Eigenvalues of S tensor ($\lambda_{x}^{2},\lambda_{y}^{2},\lambda_{z}^{2}$)&290.17 \ 2.30 \ 1.17&277.74 \ 3.54 \ 2.01&96.61 \ 69.66 \ 47.82&210.52 \ 44.45 \ 13.03\\
			Relative shape anisotropy ($\kappa$ factor)&0.99989&0.99968&0.37499&0.93135\\
		\end{tabular}
	}
	\footnotetext[1]{These values have been calculated in the area around the main peak of the distribution}
\end{table*}

Another property used to ensure convergence, in addition to the tilt-angle, is the time evolution of the non-bonding energy for the C$_{16}$S chains on every surface. The results are shown in Figure \ref{fig:energy_evolution}. The alkanethiol chains on the (111) and (211) have an excellent and very fast (below 20ns) relaxation, while the ones on the (311) converge rather slowly (above 250ns) with respect with the first two. The energy of (221) was almost constant, having however significant fluctuations. In all cases, we followed the time evolution of the systems to hundreds of ns in order to be sure that they had reached thermodynamic equilibrium prior to the calculation of any structural properties.

\begin{figure*}
	\includegraphics[width=1.0\textwidth]{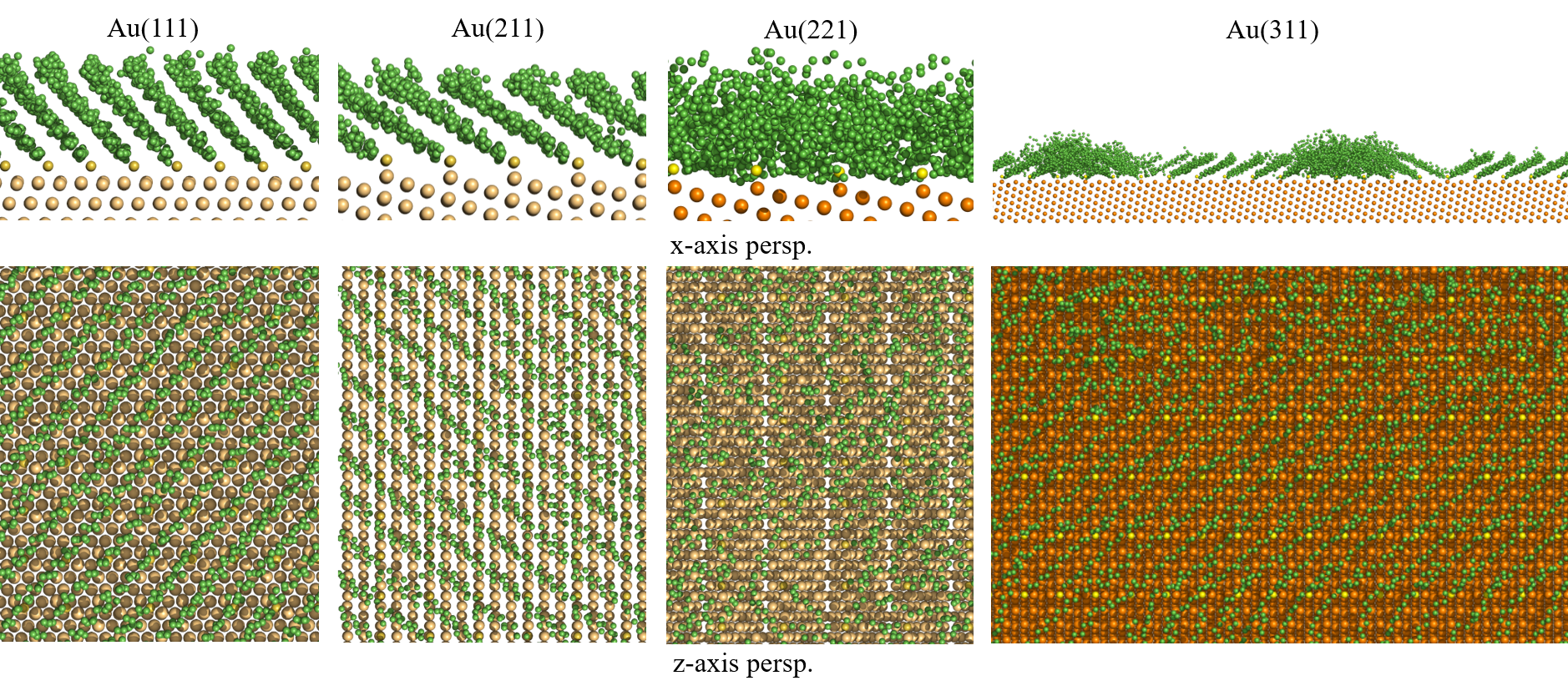}
	\caption{\label{fig:conf_fin} Final configurations of the four systems. Note the order on the (111) and (211) surfaces in comparison with the disorder on the (221) and (311) ones. Nevertheless, partially ordered formations are observed on (311).}
\end{figure*}

Typical snapshots of the final configurations of the four surfaces are shown in Figure \ref{fig:conf_fin}.
As a general observation for the final states of our systems, one can see that only two of them, (111) and (211), have reached a total order (Figure \ref{fig:conf_fin}). The (311) system was partially ordered giving large vacancies with fixed chains separated by transition zones where the alkanethiols were messed (this will be discussed later in this work). The (221) system was totally disordered. The convergence to the final state for (111), (211) and (221) systems was very fast (around 20ns), rather the one of the (311) system which seems to be slower (above 250ns) with respect to the first three.

\begin{figure}
	\includegraphics[width=0.5\textwidth]{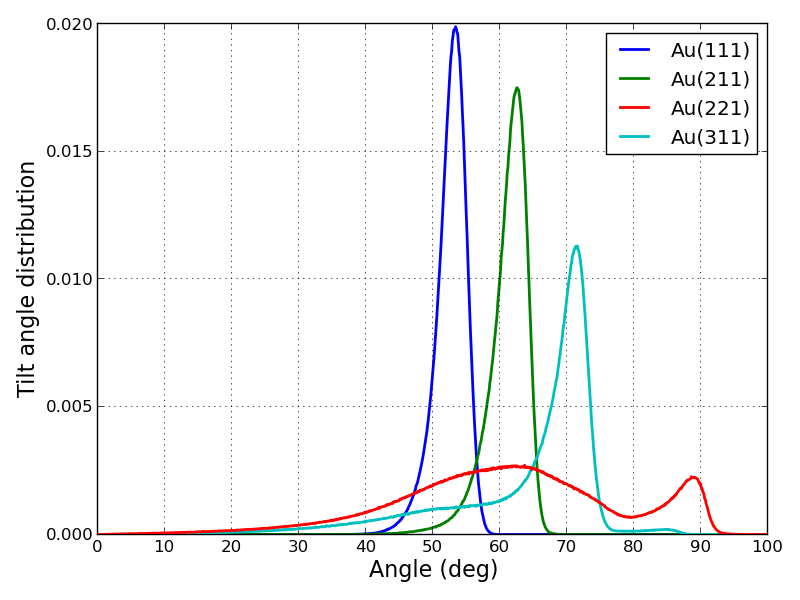}
	\caption{\label{fig:tilt_angle_distr} Tilt angles normalized distributions for the C$_{16}$S chains on various Au surfaces. While there is convergence to equilibrium for surfaces (111) and (211), its lack for (221) and (311) is evident.}
\end{figure}

The normalized distribution of tilt angles after equilibration on the examined surfaces is plotted in Figure \ref{fig:tilt_angle_distr}. From this diagram, we observe the excellent order of alkanethiols on the (111) and (211) surfaces with mean values equal to 52.6$\pm$2.8 and 61.1$\pm$3.1 degrees respectively. The result value for the (111) system seems to be analogous to the theoretical\cite{doi:10.1021/jp200447k} and experimental\cite{doi:10.1063/1.1346676,doi:10.1021/ja036143d} results mentioned before, where tilt angles lie around an average of $50^{o}$, thus confirming the method correctness.

Because of the partial order of the (311) system, its mean tilt angle was calculated in the area around the main peak as it is demonstrated in the same plot and it was found to be 69.6$\pm$3.2 degrees. The (221) system gives an average tilt angle of 61.3$\pm$16.7 degrees with a very flat distribution because of its disordered final state. However, there is a small peak near 90 degrees which indicates that there are chains almost parallel to the surface. For this system, the percentage deviation from the average is very high (27.27\%), which strengthens our view.

The fact that the tilt angle is larger on the (211) surface indicates larger interaction between the alkanethiol and this surface in comparison with the (111). Due to the shift of the distribution maxima to larger tilt angle values, Figure \ref{fig:tilt_angle_distr} also indicates increasing interaction of the chains with the (311) surface.

From Figure \ref{fig:conf_fin}, we observe ordered SAMs structures on (111), (211) and partially on (311) surfaces and disorder on the (221) as it has been mentioned above. This order and disorder can be explained considering two reasons: (a) the distances between the S atoms on the various surface positions which lead to larger distances between the $\mathrm{-CH_2-}$ and $\mathrm{-CH_3}$ groups belonging to adjacent chains and (b) the geometry of the surface structure and in particular the different step type. The (221) surface has (111) steps where one Au atom of the upper terrace is bonded to two atoms of the lower terrace. The (211) and (311) surfaces have (100) steps where one Au atom of the upper terrace is bonded to one atom of the lower terrace. Similar differences due to the surface orientation have been observed in interfaces between diamond and amorphous carbon.\cite{KOPIDAKIS20071875} Indeed, the order decreases as the grafting density of S atoms on the Au surface gets lower from (111), to (211) and (221) systems, as indicated in Table \ref{tab:char}. The partial order of the SAMs on the (311) one, can be explained by the extra step in Au surface that lies between the two adjacent S atoms (see Figure \ref{fig:Au_S_multi_total} (311)(a)) which modifies the relationship between metal-chain and chain-chain forces of the system.

\begin{figure*}
	\includegraphics[width=1\textwidth]{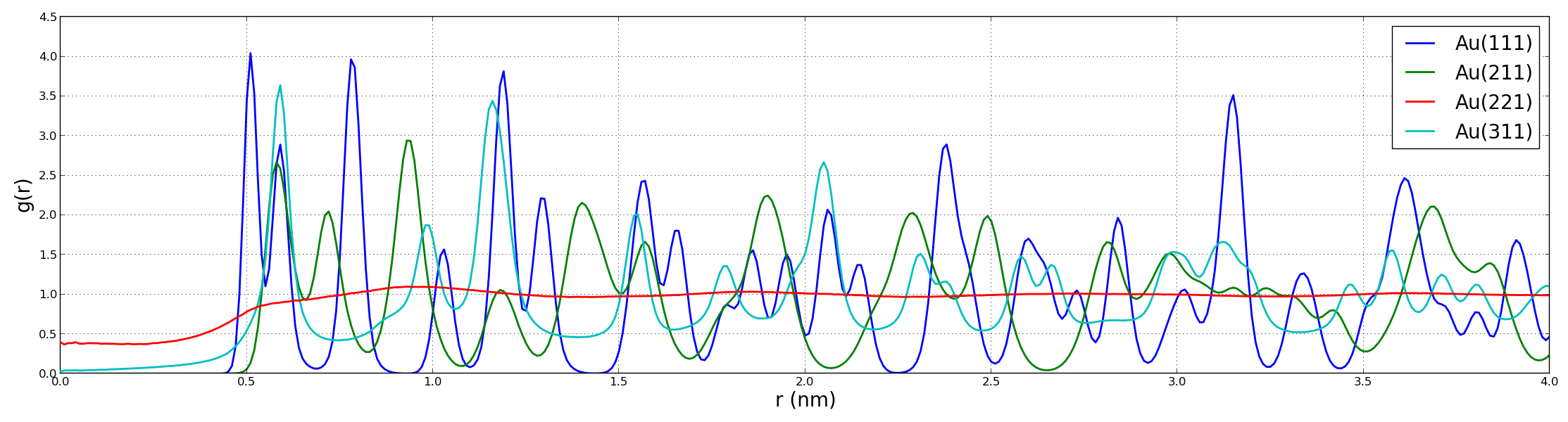}
	\caption{\label{fig:rdf_2D} Two-dimensional radial distribution function of the center of mass for groups belonging to different alkanethiol chains (intermolecular) on different Au surfaces. On (111), (211), and (311) the peaks are obvious indicating ordered structure. On the contrary, on (221) the peaks vanish, which indicates disordered configuration.}
\end{figure*}

To further explore the emerging of order or disorder depending on surface orientation, we plot the 2D radial distribution function ($g(r)$) of the center of mass of alkanethiols in Fig. \ref{fig:rdf_2D}. The peaks of $g(r)$ corresponds to distances where it is most likely to find two centers of mass.  Ordered structures show a series of distinct peaks whereas a random structure will have $g(r)=1$. As can be seen from Fig. \ref{fig:rdf_2D},  (111), (211) and (311) surfaces show clear peaks at specific distances between the center of masses of the alkane groups, while a much smoother $g(r)$ is shown for the  (221) surface. These features 2D radial distribution function suggest that alkanethiol groups on the (111), (211) and (311) systems are localized at specific positions in an ordered superstructure. This is not the case for the (221) system, where $g(r)$ has typical features for an amorphous-like or disordered system.

\subsection{\label{sec:mean_atom_height} Mean atom height and the monolayer thickness ($\mathbf{z_{tail}}$)}
\begin{figure}
	\includegraphics[width=0.5\textwidth]{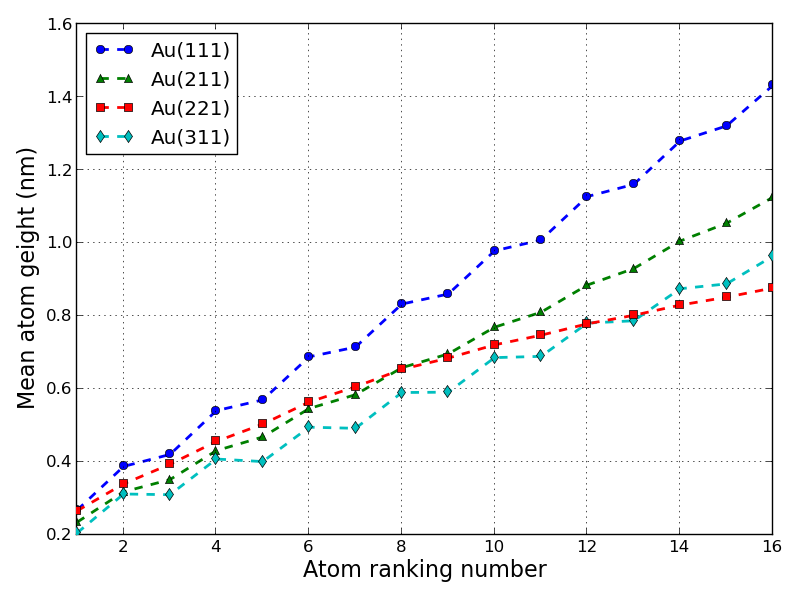}
	\caption{\label{fig:mean_atom_height} The mean atom height for every C in respect with its ranking number in the alkane chain. Linearity is obvious for the ordered configurations, while there is not exist for the rest.}
\end{figure}

The mean atomic distance of $\mathrm{C_{16}S}$ chains on the various surfaces examined in this study against the C atom ranking number are demonstrated in Figure \ref{fig:mean_atom_height}. Previous studies showed a rather linear profile of the mean distance of the alkane chains from the Au surface for ordered configurations\cite{doi:10.1021/jp067347u}; similar trends are also founded here. As this height is strongly depended on the tilt angle expressed above, one expects that a bigger value of this angle indicates a more sloping chain with respect to the vertical axis. This is true for the chains on the (111), (211) and (311) surfaces where order was observed. For the (221) surface where disorder is observed, the linearity does not exist.

The mean height of the last C atom ($z_{tail}$) is demonstrated in Table \ref{tab:results}. Similarly to the tilt angle, the mean atom height for the (111), (211) and (311) surfaces is 1.44$\pm$0.06, 1.12$\pm$0.07 and 0.96$\pm$0.22 nm respectively, while for the (221) it is 0.84$\pm$0.38 nm. The shape of the diagram for the last surface is rather a curved line than a herringbone as a result of the observed disorder.

\begin{figure}
	\includegraphics[width=0.5\textwidth]{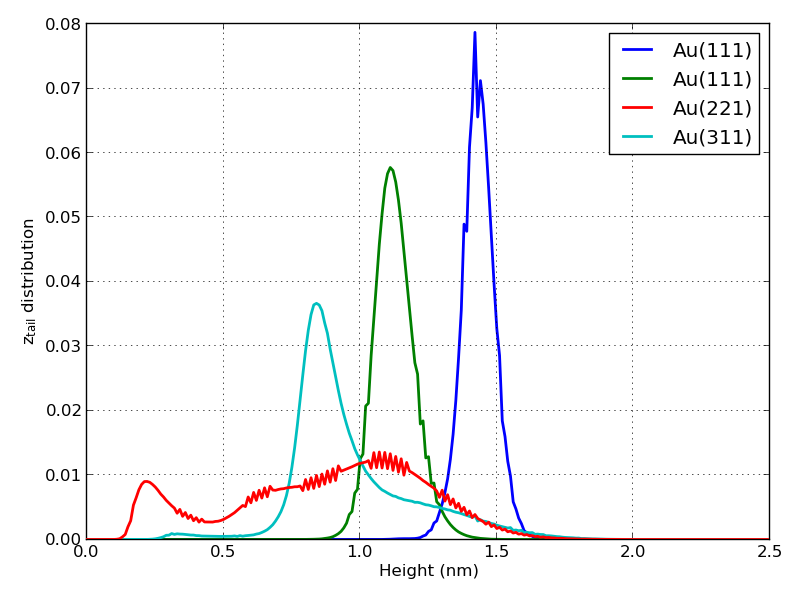}
	\caption{\label{fig:z_tail_distr} The normalized distribution of $z_{tail}$ for the last C in the all the alkane chains. Clear peaks indicate the ordered areas. Especially for the Au(311) system, the clear peak at 0.84 nm indicates the system ordered areas while the area between 1.0 and 1.6 nm indicates the non-ordered areas.}
\end{figure}

The normalized distribution of the distance of the last C atom in chain from the metal slab ($z_{tail}$) has been plotted in Figure \ref{fig:z_tail_distr}. For the (111) and (211) well-ordered systems a Gaussian-like profile is observed, with clear peaks around 1.4 and 1.1 nm, respectively, while for the (311) system with the semi-ordered behavior this peak occurs around 0.84 nm that represents $z_{tail}$ the "ordered" areas of the system. However, for the (311) system, an additional broad (slightly declined) height distribution is observed for distances between 1.0 and 1.6 nm that indicates the existence of a significant number of chains at the non-ordered areas that have some higher $z_{tail}$ than the ordered ones, but without a clear convergence into a second central value. 

Completely different is the height distribution of the last C atom for the unordered (221) system; a very broad curve is found with a flattened peak around 1.1 nm and a smaller peak around 0.2 nm. The first indicates that there is not a global average value for $z_{tail}$ for this system, while the latter shows that some chains are almost parallel to the surface.

\subsection{\label{sec:precession_angle} Precession angle ($\mathbf{\chi}$)}
\begin{figure}
	\includegraphics[width=0.5\textwidth]{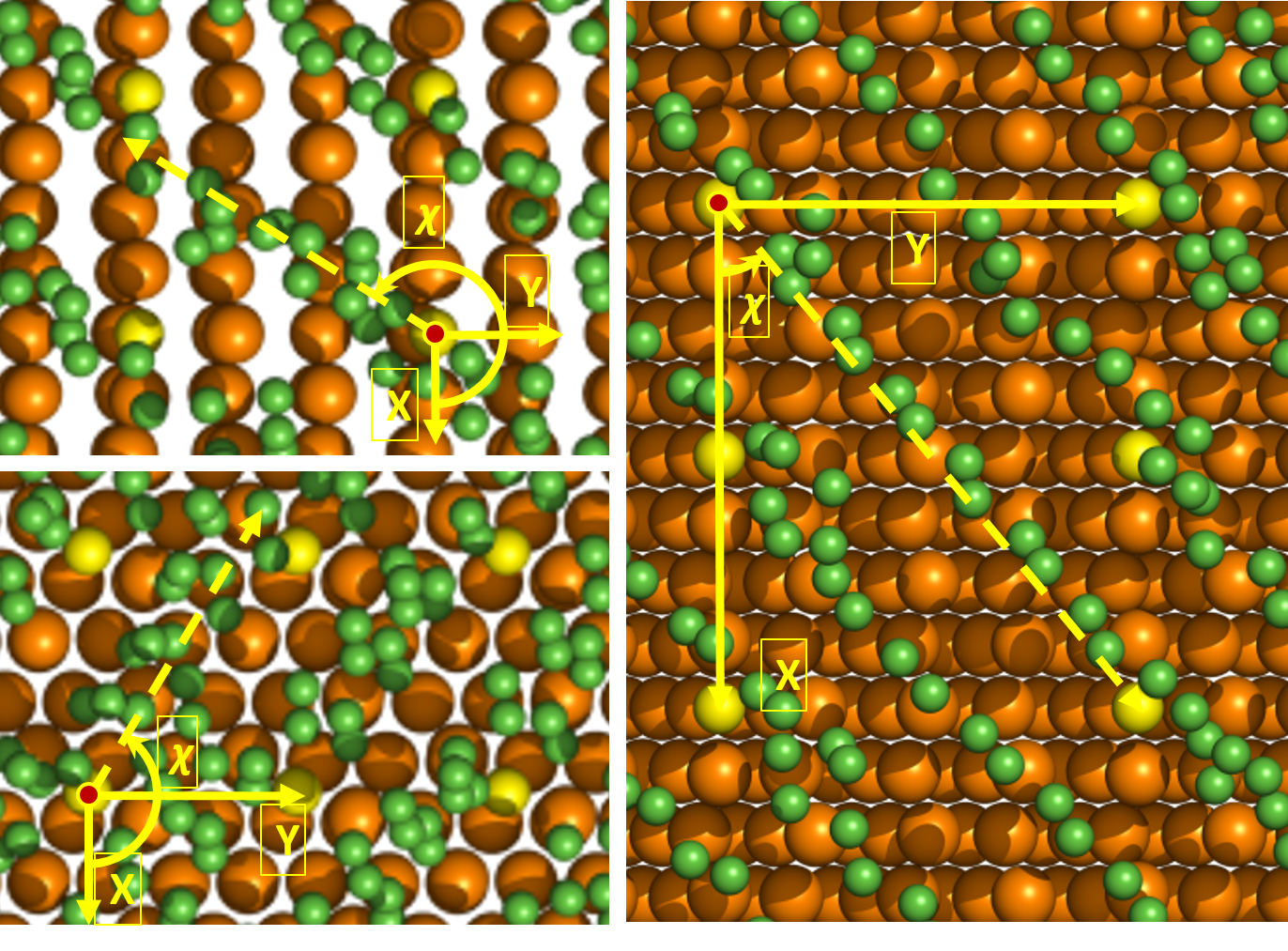}
	\caption{\label{fig:prec_angle_image} Precession angles definition for the C$_{16}$S chains on the (211) (upper left), (111) (lower left) and (311) (right) surface systems studied. The angles are measured counterclockwise from the x-direction (detail from Figure \ref{fig:conf_fin}). The axes shown correspond to the following crystallographic orientations: $x=[0\bar{1}1], y=[1\bar{1}\bar{1}]$ for (211); $x=[\bar{2}11], y=[[0\bar{1}1]]$ for (111); $x=[[0\bar{1}1]], y=[2\bar{3}\bar{3}]$ for (311).}
\end{figure}

\begin{figure}
	\includegraphics[width=0.5\textwidth]{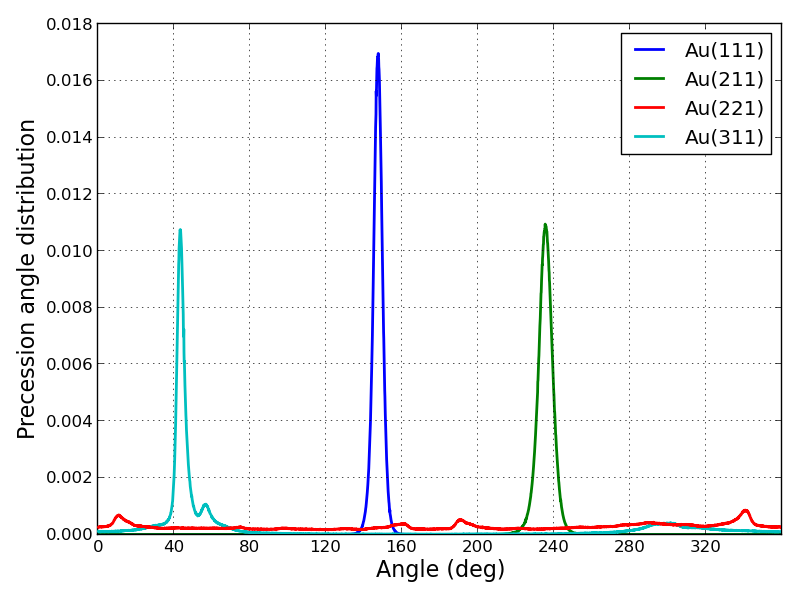}
	\caption{\label{fig:prec_angle_distr} Precession angles normalized distributions for the C$_{16}$S chains.}
\end{figure}

The precession angles of the $\mathrm{C_{16}S}$ chains are defined as shown in Figure \ref{fig:prec_angle_image} and are measured counterclockwise from x-direction. The mean values for the (111) and (211) systems are 147.2$\pm$2.7 and 235.4$\pm$4.3 degrees respectively (Table \ref{tab:results}). For the semi-ordered (311) surface this value is 43.9$\pm$2.6 degrees and has been calculated in the area around the main peak, as it has been done previously for the tilt angle as well. The values for the normalized distributions of the precession angles for all of the systems are plotted in Figure \ref{fig:prec_angle_distr}.
\\
\\
In both of the (111) and (211) ordered cases, the chains lie very close to the diagonal of the quadrilaterals formed by four neighboring S atoms. The corresponding angles for these quadrilaterals measured from the x-direction as it is shown in Figure \ref{fig:prec_angle_image} (upper left - lower left) on (111) and (211) surfaces are \textasciitilde130.9 and \textasciitilde234.7 degrees respectively, which are very close to calculated precession angles. Former studies on (111) systems with $(\sqrt{3}\times\sqrt{3})R30^{o}$ close packed  arrangement\cite{doi:10.1021/jp067347u}, have shown that the alkane axis is projected between the nearest-neighbor (NN) and the next-nearest-neighbor (NNN) of S atom that connects the alkane with the substrate, preferring an orientation towards the NNN direction. This is also true in this study for both the (111) and (211) ordered systems.

The semi-ordered (311) system indicates a different behavior: the quadrilateral mentioned above is formed by every second S atom which lies on the edge of the step (x-axis) and each S on the y-axis (Figure \ref{fig:prec_angle_image} (right)) and its diagonal forms a \textasciitilde39.7 degree angle from the x-axis according to the displayed dimensions, very close to the average precession angle. On the other hand, the (221) system, which gives a totally disordered formation, has not a distinguished peak on the plot.

\subsection{\label{sec:gauche_defects}Gauche defects}
\begin{figure}
	\includegraphics[width=0.5\textwidth]{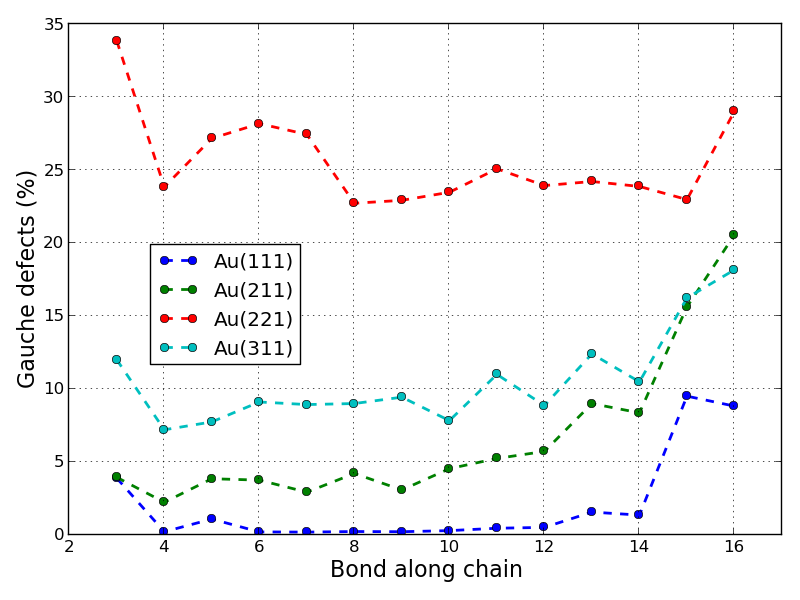}
	\caption{\label{fig:gauche} The gauche defects as they occur on the various surfaces formations.}
\end{figure}

The all-trans configuration of the C chains in the systems we study, is indicated by calculating the gauche defects percentage with respect to the bond ranking along the chains starting from the $\mathrm{C^{(1)}-C^{(2)}}$ bond (ranking number 3) until the end of the chain. The results are demonstrated in Figure \ref{fig:gauche}.

As it has been stated elsewhere\cite{doi:10.1021/jp067347u}, most "gauche defects" of ordered SAM alkanethiol chains are expected to occur in bonds far from the surface, especially in the last bond of the chain, due to more available free volume at the chain ends~\cite{VH_2002}. 	
This is clearly shown for the ordered formations on the (111), (211) (311) Au surfaces, where the percentage of gauche defects are 8.8\%, 20.6\% and 18.2\% respectively (Table \ref{tab:results}). In addition, one can observe the herringbone arrangement in the percentage of the gauche defects. Such conformations are energetically preferable in these situations because they minimize overlaps between the neighboring molecules. The observed higher values in the 3rd bond are due to the participation of the $\mathrm{C^{(2)}}$ atom in the 1st dihedral angle Au-S-$\mathrm{C^{(1)}-C^{(2)}}$ which is dominated by a different potential (see Section "Calculation of Au-S-C$^{(1)}$-C$^{(2)}$ dihedral angle potentials"\ref{sec:meth31}) and prevents the formation of bonds in trans state. Figure \ref{fig:gauche} indicates that the percentage of the gauche defects increases as the interaction of the chains with various surfaces increases. Indeed, for (111) the defects are very few, while for (211) and (311) systems, these defects are remarkably increased.

For the system on the (221) surface, where no order has been observed, the previous features seem to be fade. This is also indicated in Figure \ref{fig:gauche} by the almost random percentage values of gauche defects, while in the ``semi-ordered'' (311) final state the oscillating high values are observed.

\subsection{\label{sec:Gyration_tensor}Gyration tensor and relative shape anisotropy}
As a final measure of the structure of the C$_{16}$S/Au structures we've examined the shape of the alkanethiol chains at their final order, by calculating their radius of gyration tensor\cite{doi:10.1021/ma00148a028} that is defined by
\begin{equation}
S_{mn}\stackrel{def}{=}\frac{1}{N}\sum_{i=1}^{N}(r_{m}^{(i)}-r_{CM}^{(i)})(r_{n}^{(i)}-r_{CM}^{(i)})
\label{eq:GyrTen}
\end{equation}
where $N$ is the number of particles (here: $\mathrm{CH_x}$ united atoms) of the chain,  $r_{m}^{(i)}$ is the $m^{th}$ Cartesian coordinate of the average position vector $r^{(i)}$ of the $i^{th}$ particle and $r_{CM}^{(i)}$ is the average position vector of the center of mass of the specific chain the particle belongs to. Because of its symmetry (it is a $3\times3$ matrix), its diagonalization gives a principal axis system where we choose that
\begin{equation}
S=\mathrm{diag}(\lambda_{x}^{2},\lambda_{y}^{2},\lambda_{z}^{2})
\label{eq:eigenvals}
\end{equation}
where $\lambda_{x}^{2},\lambda_{y}^{2},\lambda_{z}^{2}$ are the eigenvalues of $S$ and $\lambda_{x}^{2}\geq\lambda_{y}^{2}\geq\lambda_{z}^{2}$.\\
The eigenvalues are demonstrated in Table \ref{tab:results}. In the three ordered  conformations there is a clear preference to a specific dimension (obvious to the chain axis dimension), as the respective eigenvalue is much greater than the other two ones. This fact can be shown further by calculating the relative shape anisotropy factor $\kappa$ as\cite{doi:10.1021/ma00148a028}
\begin{equation}
\kappa^{2}=\frac{3}{2}\frac{\lambda_{x}^{4}+\lambda_{y}^{4}+\lambda_{z}^{4}}{(\lambda_{x}^{2}+\lambda_{y}^{2}+\lambda_{z}^{2})^{2}}-\frac{1}{2}
\label{eq:k_factor}
\end{equation}
where $0\leq\kappa\leq1$. $\kappa=0$ only occurs if all particles are spherically symmetric, and $\kappa=1$ only occurs if all particles lie on a line. Indeed, the calculation of $\kappa$ (see Table \ref{tab:results}) shows that chains on the (111), (211) and (311) surfaces are very close to linearity ($\kappa$=0.99989, 0.99968 and 0.93135 respectively), while one of them (on (221) surface) is far from it ($\kappa$=0.37499).

\section{\label{sec:disc}Discussion}
In this work we investigated the structural properties and ordering of hexadecanethiol (C$_{16}$S) SAMs formed on the planar Au(111), and stepped Au(211), Au(221) and Au(311) surfaces via long detailed atomistic MD simulations. To describe accurately the interaction of C$_{16}$S chains with the Au surfaces we've extended a classical force field reported in the literature, by parametrizing the dihedral angle interaction potentials (different for each system) between Au\nobreakdash-S\nobreakdash-$\mathrm{C^{(1)}-C^{(2)}}$, where Au is the nearest surface atom to the ligant S atom of the chain and $\mathrm{C^{(1)}-C^{(2)}}$ the following C atoms. The latter potentials were calculated using DFT calculations and were described with high-degree polynomials after a fitting process. 

Comparing the morphology of the C$_{16}$S SAMs on various Au surfaces, a clear transition from well-ordered, for (111) and (211) surfaces, to ``semi-ordered'' for the (311), up to fully disordered structures for the (221) one, is observed. In particular SAMs on the (311) Au surface show regimes of ordered chains separated by non-ordered transition zones maybe as a result of the specific surface geometry. 

The structure of the C$_{16}$S SAMs systems has been quantified by calculating several different properties. The chain tilt angle is a very important property since it indicates the interaction of alkanethiol chain with the surface and the way the chains are "ordered" with respect to the planar surface. For the (111) Au surface, results are in agreement with previous theoretical and experimental data considering systems with the same coverage. We also observed that on the complex (211) and (311) surfaces the tilt angle was even larger than the one of (111), which indicates the stronger "tilting" of chains towards the Au surface for these surfaces. For the disordered (221) system a rather flat distribution of the tilt angles, between 50 and 90 degrees was observed. Consistent results were found by calculating the mean C atom heights with respect to the ranking number of each C atom  for the different Au surfaces.

The precession angle also has some interesting features: the well-ordered chains on (111) and (211) surfaces the chain axis was projected near the NNN direction, in agreement with results reported in the literature. It is very interesting that different is the case for the semi-ordered (311) system, where the ordered chain axes prefers to be projected nearly above the diagonal of the quadrilateral formed by every second S atom that lies on the edge of the surface step (x-axis) and each S atom on the vertical y-axis. This fact maybe another effect of the (311) geometry. The disordered C$_{16}$S/(221)Au system shows no preferential precession angle.

The overall morphology of C$_{16}$S chains were also studied in the intermolecular level by calculating the pair 2D radial distribution function between atoms belonging to different chains. Clear strong peaks at various distances, indicating a crystalline-like order, were found for the systems on the (111) and (211), and less for the (311), surfaces, whereas for the (221) one the peaks vanish indicating a disordered morphology.

Finally, the ordered and disordered states of the various examined systems were related with the overall shape of the C$_{16}$S chains, by calculating the gyration tensor and the relative shape anisotropy factor. For the chains on the (111), (211) and (311) Au surfaces, a value of $\kappa$ very close to 1 was found, indicating extended (almost all-trans) alkanethiol chains, whereas for the (221) system $\kappa$ is much below one.

The results reported here emphasize the role of the Au substrate on the final structure and morphology of the alkanethiol SAMs. In general, for SAMs on flat surfaces, it is expected that the structure of the SAMs would strongly depend on the grafting density, being a result of an interplay between the energetic interaction of the molecules with the surface that enhances order, and the associated entropy that leads to disorder; a transition from amorphous to ordered domains is expected as the grafting density increases (entropy decreases). Therefore, for systems with high grafting densities, as those studied here, ordered structures are to be expected. However, from our results it is clear that the geometrical characteristics of the Au substrate can also strongly affect the self-assembled structures. For the Au(111) and the stepped Au(211) the grafting densities are high (3.24 chains/nm$^{-2}$ and 2.29 chains/nm$^{-2}$ respectively) and the final structures are well ordered. For the stepped Au(331) surface (grafting density of 1.69 chains/nm$^{-2}$) domains with ordered and amorphous like chains are found. On the contrary, for the Au(221) surface, despite the fact that the grafting density is still relatively high (1.87 chains/nm$^{-2}$), clear disordered structure has been observed. This can be attributed to additional excluded volume interaction induced by the specific steps in this surface that prevent the collective arrangement of the C$_{16}$S chains in well formed structures. 

\section{\label{sec:concl}Conclusions}
The "order to disorder transition" of the C$_{16}$S chains by changing the type of the Au surface can offer a direct way to control the morphology of the SAMs by only changing the crystalline characteristics of the surface, thus providing a complementary to chemistry way to produce SAMs with the desired morphology.

The above discussion is far from leading to definite conclusions. The current work is, according to our knowledge, the first systematic theoretical/simulation study concerning the complex role of the substrate on the final properties of the SAMs. Without doubt, a lot of work needs to be done in order to examine whether the order to disorder transition observed here is seen also for other systems, and more general to clarify the role of the substrate characteristics (geometry, crystalline structure, defects, etc.) on the properties of the SAMs systems. For example, detailed studies of the structure of SAMs as a function of the grafting density on the same surface or for other metallic surfaces, such as Ag, Pt etc, are necessary to clarify whether the observations reported here are valid for other systems as well. In addition, it would be very interesting to investigate the SAMs structure if the S atoms were not frozen at their initial positions (a movement of chains might be possible, especially at low grafting densities). All these will be the subject of future works.

\section{\label{acknowledgement}Acknowledgement}
The authors thank the supporting teams of the ASE/GPAW and GROMACS open source code that contributed significantly to the successful completion of this work. They also acknowledge the CyTera, BIBLIOTHECA ALEXANDRINA (project pro17a111s1) and ARIS high-performance computing facilities for granting computing time, as well as their staff for valuable help. This work was supported by computational time granted from the National Infrastructures for Research and Technology S.A. (GRNET S.A.) in the National HPC facility - ARIS - under project IDs pr007027-NANOGOLD and pa181005-NANOCOMPDESIGN. VH acknowledges support by the project "SimEA", funded by the European Union’s Horizon 2020 research and innovation programme under grant agreement No 810660. IR, GK and DS acknowledge support from HFRI Project MULTIGOLD numbered 1303-KA10480.

\section{Bibliography}
\bibliography{mypaper}

\newpage
\section{\label{sec:suppinfo}Supplementary information}
\subsection{\label{subsec:fitting-data}Fitting Data for the calculated potential}
The calculation process of the fitting data is demonstrated in Table \ref{tab:dih_pot}. The table is divided in four sections that represent each one of the studied surfaces, as indicated. Each section is divided in three columns:
\begin{enumerate}
	\item Starting from the third column of each section, the difference between the absolute value and the lowest absolute value of calculated potential in kJ/mol is demonstrated. Thus, a zero value indicates the configuration of the lowest potential for each system.
	
	\item The first column of each section indicates the dihedral angle of Au-S-$\mathrm{C^{(1)}-C^{(2)}}$ starting from our initial configuration (0 deg). This was retrieved from data reported by Barmparis et al.[citation 18 in the paper manuscript] as the configuration of lowest adsorption energy for each surface. The fact that the initial angle generally does not correspond to the lowest energy in our systems is because we used ethanethiols instead of methanethiols as Barmparis et al. did. This caused a shift of the systems lowest energy to some adjacent angles. The only system that provided its lowest energy at the same configuration with its initial one was Au(211). The first and the last angles differ by 360 degrees and correspond of course to the same energy due to the periodicity of the potential. We notice that, in the calculation of polynomials we have ensured that the value, as well as their first and second derivatives at the initial and the final angle of calculation, are respectively equal. This was achieved with accuracy between $10^{-10}$ and $10^{-7}$ depending on the examined surface. 
	
	\item The second column of each section indicates the dihedral angle translated into the \mbox{IUPAC/IUB} convention, where the angle $\phi$ between the two planes of dihedral angle is zero at \emph{cis} position ($\phi_{rot}^{(cis)}=0$).
\end{enumerate}
Moreover, all data have been shifted properly in order to demonstrate the $\phi_{rot}^{(cis)}$ at the center of each plot (see Fig.~\ref{fig:pot_dih}).

\begin{table*}
	\caption{\label{tab:dih_pot}Data for the dihedral Au-S-$\mathrm{C^{(1)}-C^{(2)}}$ fitting for the various surfaces}
	\centering
	\resizebox{\textwidth}{!}{%
		\begin{tabular}{r|d{1}|d{-1}||r|d{1}|d{-1}||r|d{1}|d{-1}||r|d{1}|d{-1}}
			\multicolumn{3}{c||}{Au(111)}&\multicolumn{3}{c||}{Au(211)}&\multicolumn{3}{c||}{Au(221)}&\multicolumn{3}{c}{Au(311)}\\
			\hline
			\multicolumn{1}{c|}{(deg)\textsuperscript{\emph{a}}}&\multicolumn{1}{c|}{(deg)\textsuperscript{\emph{b}}}&\multicolumn{1}{c||}{(kJ/mol)\textsuperscript{\emph{c}}}&\multicolumn{1}{c|}{(deg)\textsuperscript{\emph{a}}}&\multicolumn{1}{c|}{(deg)\textsuperscript{\emph{b}}}&\multicolumn{1}{c||}{(kJ/mol)\textsuperscript{\emph{c}}}&\multicolumn{1}{c|}{(deg)\textsuperscript{\emph{a}}}&\multicolumn{1}{c|}{(deg)\textsuperscript{\emph{b}}}&\multicolumn{1}{c||}{(kJ/mol)\textsuperscript{\emph{c}}}&\multicolumn{1}{c|}{(deg)\textsuperscript{\emph{a}}}&\multicolumn{1}{c|}{(deg)\textsuperscript{\emph{b}}}&\multicolumn{1}{c}{(kJ/mol)\textsuperscript{\emph{c}}}\\
			\hline
			40  & -176.9 & 0        & -10 & -176.5 & 0.50053  & 10  & -181.1 & 0.22168  & -100 & -177.6 & 0.98058  \\
			50  & -166.9 & 1.92209  & 0   & -166.5 & 0        & 20  & -171.1 & 1.41083  & -90  & -167.6 & 2.0242   \\
			60  & -156.9 & 4.12734  & 10  & -156.5 & 2.9621   & 30  & -161.1 & 2.81249  & -80  & -157.6 & 3.42303  \\
			70  & -146.9 & 7.53486  & 20  & -146.5 & 2.08739  & 40  & -151.1 & 4.06308  & -70  & -147.6 & 4.27523  \\
			80  & -136.9 & 10.84462 & 30  & -136.5 & 5.32198  & 50  & -141.1 & 5.44702  & -60  & -137.6 & 4.54192  \\
			90  & -126.9 & 13.1409  & 40  & -126.5 & 9.18535  & 60  & -131.1 & 4.75483  & -50  & -127.6 & 4.34841  \\
			100 & -116.9 & 14.57073 & 50  & -116.5 & 11.98105 & 70  & -121.1 & 5.02124  & -40  & -117.6 & 3.55137  \\
			110 & -106.9 & 17.17039 & 60  & -106.5 & 15.99277 & 80  & -111.1 & 4.59373  & -30  & -107.6 & 2.24303  \\
			120 & -96.9  & 20.64851 & 70  & -96.5  & 17.38566 & 90  & -101.1 & 4.17872  & -20  & -97.6  & 1.02032  \\
			130 & -86.9  & 22.73943 & 80  & -86.5  & 22.54092 & 100 & -91.1  & 2.9578   & -10  & -87.6  & 0.37624  \\
			140 & -76.9  & 22.88312 & 90  & -76.5  & 20.73711 & 110 & -81.1  & 0        & 0    & -77.6  & 0.1412   \\
			150 & -66.9  & 23.254   & 100 & -66.5  & 17.31129 & 120 & -71.1  & 0.83306  & 10   & -67.6  & 0.99838  \\
			160 & -56.9  & 22.92139 & 110 & -56.5  & 18.7662  & 130 & -61.1  & 1.0445   & 20   & -57.6  & 2.33373  \\
			170 & -46.9  & 22.81019 & 120 & -46.5  & 18.76449 & 140 & -51.1  & 1.50495  & 30   & -47.6  & 5.25309  \\
			180 & -36.9  & 23.06272 & 130 & -36.5  & 20.18011 & 150 & -41.1  & 5.093    & 40   & -37.6  & 7.38145  \\
			190 & -26.9  & 23.22129 & 140 & -26.5  & 19.58138 & 160 & -31.1  & 5.9225   & 50   & -27.6  & 10.69598 \\
			200 & -16.9  & 21.75202 & 150 & -16.5  & 20.05389 & 170 & -21.1  & 9.33662  & 60   & -17.6  & 14.22945 \\
			210 & -6.9   & 21.50898 & 160 & -6.5   & 20.95312 & 180 & -11.1  & 12.62207 & 70   & -7.6   & 16.7203  \\
			220 & 3.1    & 20.04005 & 170 & 3.5    & 18.82854 & 190 & -1.1   & 15.91302 & 80   & 2.4    & 19.02913 \\
			230 & 13.1   & 18.50947 & 180 & 13.5   & 19.6641  & 200 & 8.9    & 15.4219  & 90   & 12.4   & 21.00888 \\
			240 & 23.1   & 16.904   & 190 & 23.5   & 16.377   & 210 & 18.9   & 20.54621 & 100  & 22.4   & 22.58012 \\
			250 & 33.1   & 13.12848 & 200 & 33.5   & 9.82463  & 220 & 28.9   & 17.22396 & 110  & 32.4   & 23.46611 \\
			260 & 43.1   & 10.1456  & 210 & 43.5   & 7.72552  & 230 & 38.9   & 15.69446 & 120  & 42.4   & 24.34837 \\
			270 & 53.1   & 5.85983  & 220 & 53.5   & 4.39806  & 240 & 48.9   & 15.44822 & 130  & 52.4   & 24.77496 \\
			280 & 63.1   & 3.15267  & 230 & 63.5   & 4.733    & 250 & 58.9   & 16.2159  & 140  & 62.4   & 24.10734 \\
			290 & 73.1   & 1.53332  & 240 & 73.5   & 1.15543  & 260 & 68.9   & 20.20456 & 150  & 72.4   & 22.64752 \\
			300 & 83.1   & 0.89677  & 250 & 83.5   & 1.53741  & 270 & 78.9   & 17.38059 & 160  & 82.4   & 21.13333 \\
			310 & 93.1   & 0.56707  & 260 & 93.5   & 2.71241  & 280 & 88.9   & 16.19414 & 170  & 92.4   & 19.28863 \\
			320 & 103.1  & 1.33393  & 270 & 103.5  & 3.21506  & 290 & 98.9   & 17.62774 & 180  & 102.4  & 16.69052 \\
			330 & 113.1  & 1.84689  & 280 & 113.5  & 4.60673  & 300 & 108.9  & 15.50676 & 190  & 112.4  & 12.59822 \\
			340 & 123.1  & 1.83621  & 290 & 123.5  & 5.94061  & 310 & 118.9  & 10.67734 & 200  & 122.4  & 8.59163  \\
			350 & 133.1  & 1.91146  & 300 & 133.5  & 6.52637  & 320 & 128.9  & 6.9842   & 210  & 132.4  & 5.27355  \\
			0   & 143.1  & 2.20951  & 310 & 143.5  & 6.581    & 330 & 138.9  & 5.61106  & 220  & 142.4  & 2.88328  \\
			10  & 153.1  & 1.62349  & 320 & 153.5  & 4.94211  & 340 & 148.9  & 1.8958   & 230  & 152.4  & 1.17682  \\
			20  & 163.1  & 0.66892  & 330 & 163.5  & 3.73802  & 350 & 158.9  & 3.37958  & 240  & 162.4  & 0.17125  \\
			30  & 173.1  & 0.04899  & 340 & 173.5  & 3.21287  & 0   & 168.9  & 0.67759  & 250  & 172.4  & 0        \\			
			40  & 183.1  & 0        & 350 & 183.5  & 0.50053  & 10  & 178.9  & 0.22168  & 260  & 182.4  & 0.98058  \\
		\end{tabular}
	}\\
	\textsuperscript{\emph{a}}Angle from the original position.\\
	\textsuperscript{\emph{b}}Angle with respect to $\phi_{rot}^{(cis)}=0$, according the \mbox{IUPAC/IUB} convention.\\
	\textsuperscript{\emph{c}}Potential difference bettween the calculated value and the lowest calculated value.\\
\end{table*}

\subsection{\label{subsec:pot_vs_dih}Fitting plots for calculated potential vs. $\mathrm{Au-S-C^{(1)}-C^{(2)}}$ dihedral angle}
\begin{figure*}
	\includegraphics[width=1.0\textwidth]{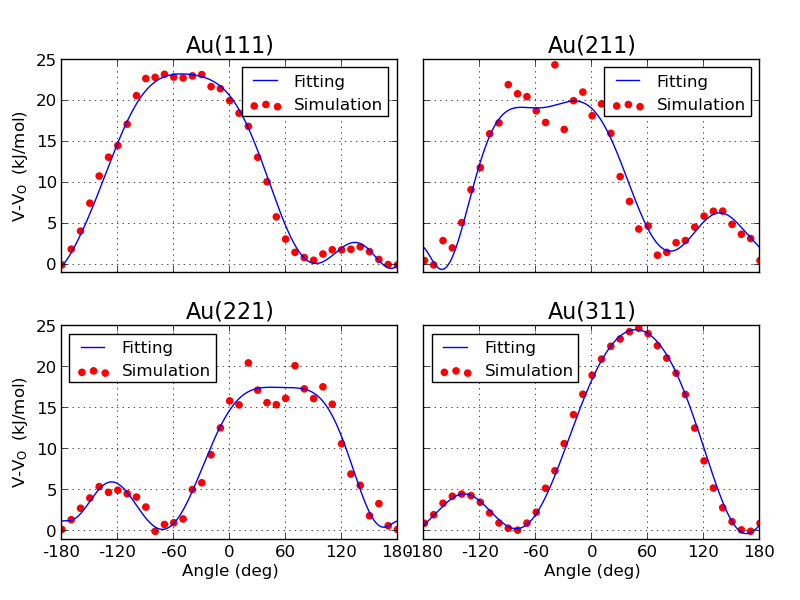}
	\caption{\label{fig:pot_dih} Potential vs. the Au-S-$\mathrm{C^{(1)}-C^{(2)}}$ dihedral angle on Au (211) and (311) surfaces. \emph{Red:} simulation data, \emph{Blue}: fitting data. Fitting is not so good in sites where there was strong repulsion to C atoms from the slab atoms. Angles with respect to $\phi_{rot}^{(cis)}=0$, according the \mbox{IUPAC/IUB} convention.}
\end{figure*}

The fitting plots of the four calculated potentials described above are demonstrated in Fig.~\ref{fig:pot_dih}. Fitting data are demonstrated in Table \ref{tab:dih_pot}. The plots for Au(111) and Au(311) fit in the calculated data points quite well, while for the rest of them (Au(211) and Au(221)) the fitting is not so good at sites near to 0 degrees ($\phi_{rot}^{(cis)}=0$) where the second C atom seems to ``penetrate'' into the surface. The reason is that these are non permitted sites because the energy of the system is very high there. However, in spite of the deviation from the expected accuracy, we consider that these potential functions fit quite well the purpose they were constructed for.\\

\subsection{\label{subsec:sq_factor}2D structure factor, S(q)}
We extracted the structure factor, $S(q)$, for all the examined systems from data of the radial distribution function shown in Fig. 8 of the text. Given the pair correlation function, $g(r)$, in 2D, the structure factor is calculated using the formula  $$S(q) = 1 + 2\pi\rho \int{g(r)\frac{\sin{qr}}{qr}rdr}$$ or by its ``discretized'' form: $$S(q_{k}) = 1 + 2\pi\rho \sum_{i=0}^{n-1}g(r_{i})\frac{\sin{q_{k}r_{i}}}{q_{k}r_{i}}r_{i}\Delta r$$
where: $n~=~900$ is the number of different distances between two chain CMs, $\rho$ is the area density of the CMs, $r_{i}$ is the distance between any two chain CMs, $\Delta r~=~0.01~nm$ is the elementary step of the distance, $g(r_{i})$ the value of the radial distribution function for that distance and $q_{k}=\frac{2\pi}{r_{k}}$.

\begin{figure}
	\includegraphics[width=0.5\textwidth]{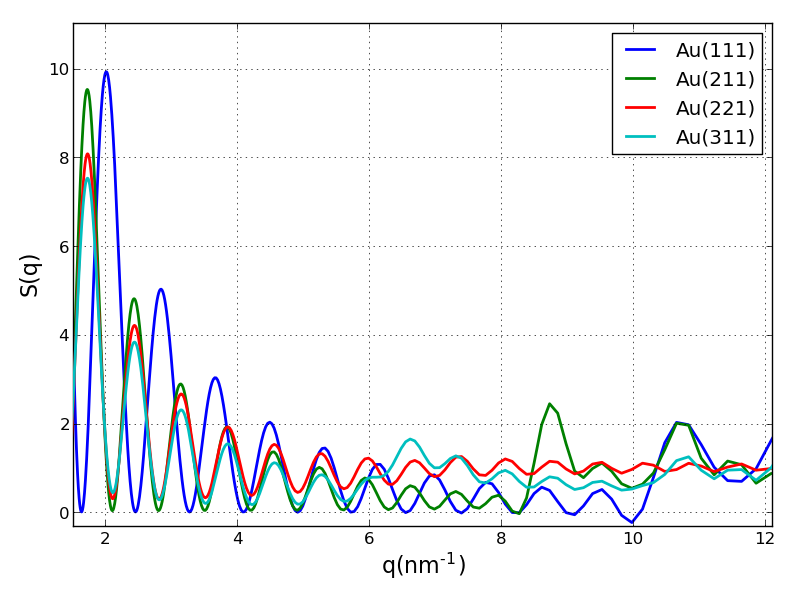}
	\caption{\label{fig:sq_2D} Two-dimensional structure factror for the SAM systems considered in the present study.}
\end{figure}

The structure factors for the four system are plotted  in Figure \ref{fig:sq_2D} where we plotted $S(q)$ for $q$ up to 12$\mathrm{nm^{-1}}$. Comparing this plot with Fig. 8 of the paper, one can observe that there are peaks of $q$'s (1 - 12 $\mathrm{nm^{-1}}$) corresponding to distances 0.5 up to 6 nm between two CMs of the chains indicating the order of both (111) and (211) systems. This is also true for the (311) despite its semi-ordered configuration. On the contrary, in the (221) unordered systems there are not such clear peaks which is what we expected because of the system's CMs aperiodicity.

\end{document}